\documentclass[12pt]{iopart}
\usepackage{graphicx,latexsym}
\usepackage{dcolumn,url}
\usepackage{iopams}
\pdfoutput=1
\begin{document}

%-----------------------------------------------------------------
\title[Time-dependent transport via the generalized master equation]
      {Time-dependent transport via the generalized master equation
       through a finite quantum wire\\ with an embedded subsystem}

\author{Vidar Gudmundsson$^{1,5}$, Cosmin Gainar$^1$,
        Chi-Shung Tang$^2$, Valeriu Moldoveanu$^3$, and Andrei Manolescu$^4$}
\address{$^1$Science Institute, University of Iceland, Dunhaga 3,
        IS-107 Reykjavik, Iceland\\
        $^2$Department of Mechanical Engineering,
        National United University,
        1, Lienda, Miaoli 36003, Taiwan\\
        $^3$National Institute of Materials Physics, P.O.\ Box MG-7,
        Bucharest-Magurele, Romania\\
        $^4$Reykjavik University, School of Science and Engineering,
        Kringlan 1, IS-103 Reykjavik, Iceland\\
        $^5$Physics Division, National Center for Theoretical Sciences,\\
        P.O.\ Box 2-131, Hsinchu 30013, Taiwan}
\ead{\mailto{vidar@raunvis.hi.is},\mailto{cstang@nuu.edu.tw}}
%
%
%----------------------------------------------------------------

\begin{abstract}
The authors apply the generalized master equation to analyze time-dependent transport
through a finite quantum wire with an embedded subsystem. The parabolic
quantum wire and the leads with several subbands are described by a
continuous model. We use an approach originally developed for a tight-binding
description selecting the relevant states for transport around the
bias-window defined around the values of the chemical potential in the
left and right leads in order to capture the effects of the nontrivial geometry
of the system in the transport. We observe a partial current reflection as
a manifestation of a quasi-bound state in an embedded well and the formation of
a resonance state between an off-set potential hill and the boundary of the
system.
\end{abstract}


\maketitle

%
%------------------------------------------------------
%

% Introduction
%---------------------
\section{Introduction}

Studying the dynamics of a few-level quantum system in contact to its environment
is an old problem in statistical mechanics and solid state physics.
The typical example comes from quantum optics where an atom is exposed to
electromagnetic radiation described by a quantum or semiclassical bosonic
bath (reservoir) \cite{Scully97}. The natural theoretical tool for investigating
transitions and computing life times is the reduced density operator (RDO) method which provides
dynamical information about the atomic system while averaging out the reservoir degrees of
freedom \cite{Carmichael99}. The time-evolution of the occupation probabilities
is given by the so-called master equation. When writing the equation of motion for the off-diagonal
matrix elements of the density operator one ends up with a generalized form of the master equation
whose mathematical structure was investigated by many authors, (see e.g.\
\cite{Haake73:98,Lindblad76:119} and the references therein).

As the system-reservoir picture is rather general the same method can be used to investigate electronic
transport at the nanoscale: one considers a mesoscopic structure which opens to particle reservoirs (leads)
and looks for the electronic flow through the nanosystem in the presence of a bias and/or time-dependent fields.
In this setup the contacts between the leads and the sample play the role of the system-reservoir coupling
Hamiltonian. Since the reduced density operator method focuses on the dynamics of the sample it can be
used to describe the transient regime, pump-and-probe experiments \cite{Fujisawa03:R1395} and
counting statistics \cite{Gustavsson06:076605}.
Theoretical calculations for few-level quantum dots have been performed
by several authors \cite{Harbola06:235309,Nyvold05:195330,Li05:205304}.
Bruder and Sch\"{o}ller \cite{Bruder94:1076,Bruder94:18436} solved a quantum master
equation for the diagonal elements of the density matrix, while K\"{o}nig {\it et al.}\
\cite{Koenig96:16820} developed a real-time diagrammatic technique for the reduced
density operator.
Usually one assumes that the memory effects can be neglected and looks for steady-state
currents within the Born-Markov and rotating wave approximation (RWA). In particular in the RWA the diagonal
and off-diagonal elements of the reduced density operator are decoupled \cite{Harbola06:235309}.
Moreover, in the high bias limit the generalized master equation reacheas the simpler form obtained previously
by Gurvitz \cite{Gurvitz96:15932}.
Non-Markovian effects were also considered in more recent
works \cite{Kleinekathofer04:2505,Welack06:044712,Vaz08:012012,Braggio06:026805}.
In particular Vaz {\it et al.}\ \cite{Vaz08:012012} took
advantage of the Laplace transform and wrote down the Redfield tensor for a two-level system. Obviously,
such a task becomes too difficult for a more complex systems. A scheme including interaction effects
was recently presented by Welack {\it et al.}\ \cite{Welack08:195315}.
Various ways to solve the generalized master equation were discussed and compared in the recent paper of
Timm \cite{Timm08:195416}.

In the references mentioned above one deals with rather simple two-level systems and their spectral
properties do not follow from a specific geometry. Also, for very small quantum dots the precise location
of the contacts is not important so it is reasonable to consider a tunneling Hamiltonian that does not contain
information about the localization of the coupled states.
In our recent paper \cite{Moldoveanu2008:GME} we have solved a non-Markovian generalized master equation (GME)
for many-level systems described within a lattice model, paying special attention to the
geometry of the system. More precisely, under a strong perpendicular magnetic field the
energy spectrum of the system is a Hofstadter spectrum with edge and bulk states. 
Given the very different nature of the sample states one expects them to carry different 
currents and (in the transient regime) even with a clear dependence on the contact region. 
We have therefore explicitly constructed a tunneling Hamiltonian
that takes care of the location of the contacts and also depends on energy. 
We have analyzed transient currents and their dependence on various parameters 
of the system as well as on the initial configuration.

The aim of this paper is to implement the same method for continuous models and to focus on the internal
dynamics of the system. As an application we consider both a pure finite quantum wire with
parabolic confinement and a wire with an embedded subsystem (a Gaussian well or potential barrier).
The finite wire is connected to semi-infinite leads of the same width. We attempt to identify
effects due to the underlying subband structure and also the formation of bound states due to the
presence of the embedded potentials. Another motivation of this work is to compare the results
of the present method with the ones obtained previously via the time-dependent Lippmann-Schwinger
formalism \cite{Thorgilsson07:0708.0103,Gudmundsson08:035329}.
Although one expects serious technical problems in the continuous model due to
the large number of states and the quite complex form of the tunneling term we show here that
one can actually say a lot about the time-dependent transport in extended systems by selecting a set of
single particle states that are expected to be relevant for the transport process.

The paper is organized as follows: Section II presents the model and the main equations, Section III
contains the numerical simulations and their discussions while Section IV summarises the results.
Some technical details are left to Appendix A.

%
% Model
%---------------------------------
\section{Model}
We consider a two-dimensional wire in the $xy$-plane.
In the $y$-direction the electrons in it are
parabolically confined with the characteristic energy $\hbar\Omega_\mathrm{0}$,
but in the $x$-direction they are confined by hard walls at $x=\pm L_x/2$.
The corresponding single-electron Hamiltonian is
\begin{equation}
       h_\mathrm{S}=h_\mathrm{0}+V={\frac{\mathbf{p}^2}{2m^*}}
       +\frac{1}{2} m^*{\Omega^2_\mathrm{0}} y^2 + V(\mathbf{r}),
\label{h_S}
\end{equation}
where $V(\mathbf{r})$ is a potential representing a subsystem embedded in the wire.
The eigenfunctions of $h_\mathrm{S}$ are denoted by $\Psi_a^\mathrm{S}(\mathbf{r})$,
with $a$ referring to the two quantum numbers of the system, $n_x^\mathrm{S}$, and
$n_y^\mathrm{S}$. The eigenvalues of $h_\mathrm{S}$ are denoted by $E_a$.
Semi-infinite leads of same parabolic confinement are attached
to the finite wire or the system at $x=\pm L_x/2$.
The single-electron Hamiltonian of the left and the right leads
is noted by $h_\mathrm{L}$ or $h_\mathrm{R}$, respectively. Their eigenfunctions are
$\Psi_q^{\mathrm{L,R}}(\mathbf{r})$ and the eigenvalues are $\epsilon^{L,R}(q)$.
Due to the subband structure of the leads
the quantum number $q$ stands both for the continuous wavenumber
and the subband index $n_y^{\mathrm{L,R}}$. The semi-infinite leads have a hard
wall at $x=\pm L_x/2$, but at time $t=t_0$ the system is opened by coupling
it to the leads in a time dependent fashion described below.
Further information about the single electron states of the quantum wire
and the leads is in Appendix A. The parabolic confinement of the system and the
leads brings with it a natural lengthscale $a_w$ with $a_w^2\Omega_0=\hbar /m^*$.

Using the eigenfunctions introduced above one can write the Hamiltonians
of the disconnected subsystems in the spectral representation
\begin{equation}
      h_S=\sum_a E_a|\Psi_a^\mathrm{S}\rangle\langle \Psi_a^\mathrm{S}|,\hspace*{0.1cm}
      h_{L,R}=\sum_q \epsilon^{L,R}(q) |\Psi^l_q\rangle \langle\Psi^l_q |.
\label{h_hat}
\end{equation}
In order to describe the coupling between the two subsystems we shall add an
off-diagonal pertubation to $h_S+h_{L}+h_{R}$
\begin{equation}
      h_T(t)=\sum_{l=L,R}\sum_a\sum_q \chi^l(t)(T^l_{qa}|
      \Psi^\mathrm{S}_a\rangle\langle\Psi^l_q|+h.c),
\end{equation}
where the coefficients $T^l_{qa}$ are meant to describe a coupling between pairs of
eigenstates $\{\Psi^l_q,\Psi^{\mathrm S}_a\}$ and will be defined below.
The time-dependent part of the coupling is regulated by a switching function
$\chi^\mathrm{L,R}(t)$.

\begin{figure}[htbq]
      \begin{center}
      \includegraphics[width=0.52\textwidth,angle=0]{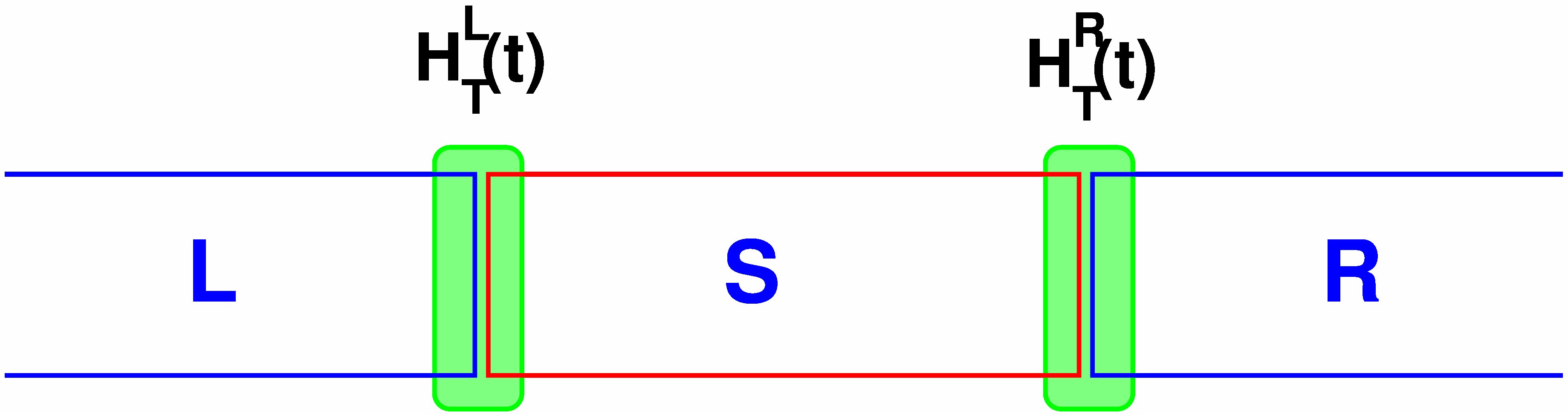}
      \end{center}
      \caption{A schematic view of the coupling of the system
               to the leads. The green shaded areas correspond to the contact regions
               defined by the nonlocal overlap function $g^{\mathrm{L,R}}_{aq}$ in $H_T(t)$.}
      \label{System}
\end{figure}

\begin{figure}[htbq]
      \begin{center}
      \includegraphics[width=0.32\textwidth,angle=0]{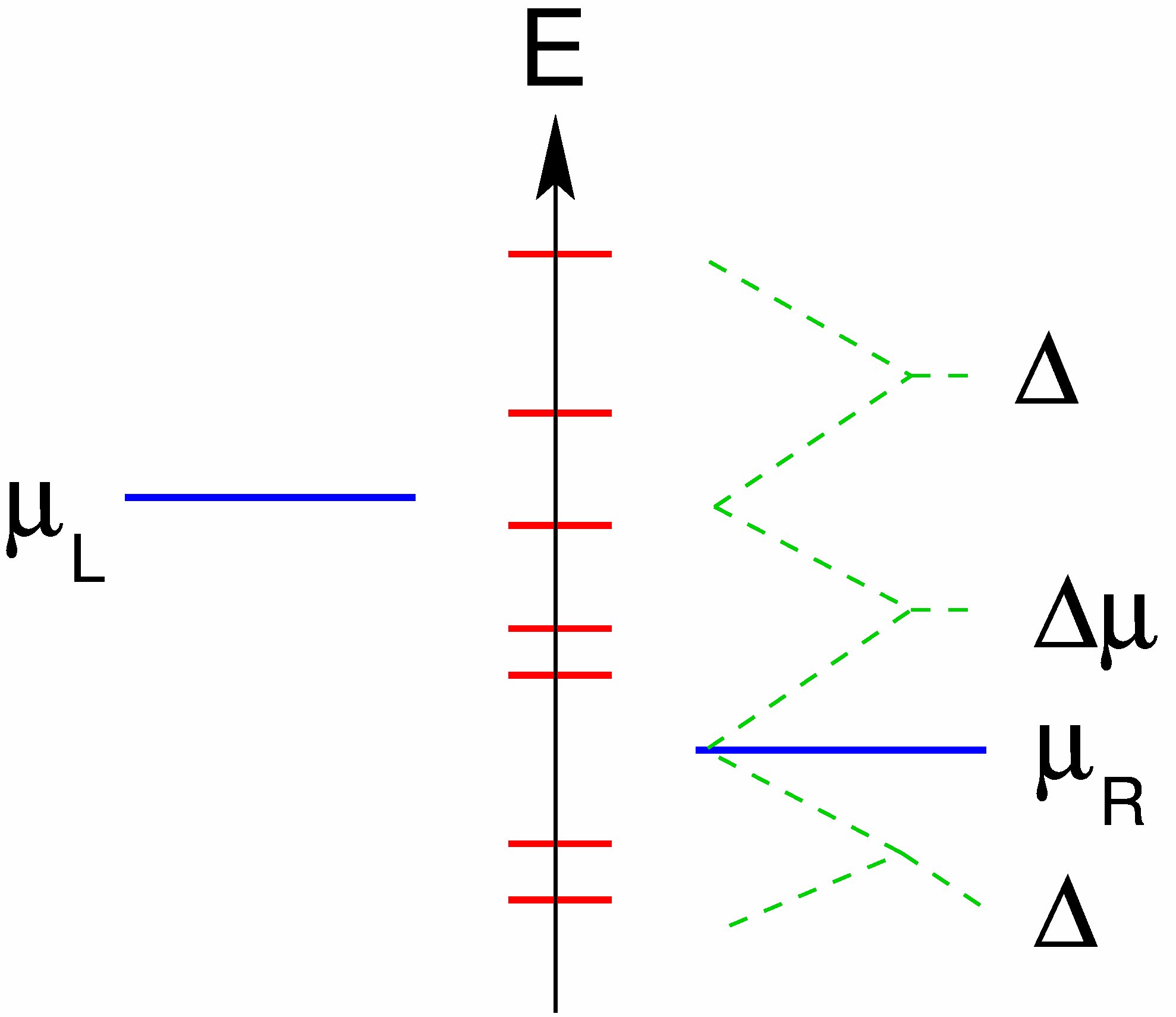}
      \end{center}
      \caption{A schematic view of the chemical potential
               bias of the leads and the relevant single-electron states
               of the system (Red). $\Delta\mu$ is the bias window.}
      \label{MuWin}
\end{figure}

In light of the variable number of electrons in the open
system as a function of time and their correlations caused 
by the coupling to the leads we resort to a many-electron description introducing creation
(annihilation) operators for electrons in the leads $c^\dagger_{ql}$ ($c_{ql}$)
and in the system $d^\dagger_{a}$ ($d_{a}$), with $l=\mathrm{L,R}$.
The many-electron Hamiltonian of the system and the leads is then
\begin{equation}
      H(t)=\sum_a E_a d^\dagger_a d_a
      +\sum_{q,l=\mathrm{L,R}} \epsilon^l(q) c^\dagger_{ql} c_{ql}
      +H_\mathrm{T}(t),
\end{equation}
where the standard tunneling Hamiltonian $H_\mathrm{T}(t)=H^\mathrm{L}_\mathrm{T}(t)+H^\mathrm{R}_\mathrm{T}(t)$
describes the coupling of the system to the left and right leads
\begin{equation}\label{H_tun}
      H^l_\mathrm{T}(t)=\chi^l(t)\sum_{q,a}
      \left\{T^l_{qa}c_{ql}^\dagger d_a + (T^l_{qa})^*d_a^\dagger c_{ql} \right\}.
\end{equation}
In our previous work Ref.\ (\cite{Moldoveanu2008:GME}) where the system
is described on a lattice we have explicitly
constructed the coefficients $T^l_{qa}$ for a neareast neighbor coupling appropriate for the
lattice model. More precisely, by denoting by $0$ the site of the one-dimensional
lead $l$ which is to be coupled to its neighbor site $i_l$ of the sample $T^l_{qa}$ was
introduced as follows
\begin{equation}\label{T_qalat}
      T^l_{qa}\propto (\Psi_q^l(0))^*\Psi^\mathrm{S}_a(i_l).
\end{equation}
Let us now construct an appropriate generalization of Eq.\ (\ref{T_qalat})
in the context of the continuous model under study.
Basically, since we need the states of the subsystems to be mutually orthogonal
in order to derive the GME we shall integrate a non-local overlap of a pair of
eigenstates $(\Psi^\mathrm{S}_a,\Psi_q^l)$ on a domain
$\Omega_S^l\times \Omega_l$ defining the contact between the sample and the $l$-th lead.
%In order to account for the geometrical structure of the system
The coupling strength tensor $T^l_{qa}$ is modeled as
\begin{equation}
      T^l_{aq} = \int_{\Omega_S^l\times \Omega_l} d{\bf r}d{\bf r}'
      \left(\Psi^l_q ({\bf r}') \right)^*\Psi^S_a({\bf r})
      g^l_{aq} ({\bf r},{\bf r'})+h.c.
\label{T_aq}
\end{equation}
The integration domains for the leads are chosen to be
\begin{eqnarray}
      \Omega_\mathrm{L}&=&\left\{ (x,y)|\left[-\frac{L_x}{2}-2a_w,
                 -\frac{L_x}{2}\right]\times [-3a_w,+3a_w]\right\}, \nonumber\\
      \Omega_\mathrm{R}&=&\left\{ (x,y)|\left[+\frac{L_x}{2},
                 +\frac{L_x}{2}+2a_w\right]\times [-3a_w,+3a_w]\right\},
\end{eqnarray}
and for the system
\begin{eqnarray}
      \Omega_\mathrm{S}^\mathrm{L}&=&\left\{ (x,y)|\left[-\frac{L_x}{2},
      -\frac{L_x}{2}+2a_w\right]\times [-3a_w,+3a_w]\right\}, \nonumber\\
      \Omega_\mathrm{S}^\mathrm{R}&=&\left\{ (x,y)|\left[+\frac{L_x}{2}-2a_w,
      +\frac{L_x}{2}\right]\times [-3a_w,+3a_w]\right\}.
\end{eqnarray}
The function
\begin{equation}
      g^l_{aq} ({\bf r},{\bf r'}) =
                   g_0^l\exp{\left[-\delta_1^l(x-x')^2-\delta_2^l(y-y')^2\right]}
                   \exp{\left(\frac{-|E_a-\epsilon^l(q)|}{\Delta_E^l}\right)}.
\label{gl}
\end{equation}
with ${\bf r}\in\Omega_\mathrm{S}^l$ and ${\bf r}'\in\Omega_l$
defines the coupling of any two single-electron states by the `nonlocal overlap'
of their wave functions in the contact region of the leads and the system, and their affinity
in energy. A schematic view of the coupling is presented in Fig.\ \ref{System}. The parameters
$\delta_1^l$ and $\delta_1^l$ define the spatial range of the coupling within the chosen
domains $\Omega_\mathrm{S}^l\times\Omega_l$.
We will be analyzing systems with complex subband structure and
geometry. For that purpose we need quite many single electron states (SESs) that in turn
lead to a requirement of an exponential number of many-electron states (MESs).
In order to deal with this computational problem we define a
window of relevant SESs with energies in the interval $[\mu_R-\Delta ,\mu_L+\Delta]$
that will consequently be used to build the
MESs used in the calculation, see Fig.\ \ref{MuWin}.
Whether, the window of SESs is large enough in a specific calculation can
only be decided by numerical experimention. Below this window all states are
considered permanently occupied and above it all states remain empty
(see Ref.\ \cite{Moldoveanu2008:GME}).

As has been described in detail earlier \cite{Moldoveanu2008:GME} we shall
deploy an occupation number basis constructed from the SESs of the isolated
finite quantum wire $\{\Psi_a^\mathrm{S}\}$. The MES $\mu$ is then
\begin{equation}
      |\mu\rangle = |i^\mu_1,i^\mu_2,\dots,i^\mu_n,\dots\rangle ,
\end{equation}
where the number $i^\mu_n$ is 1 if the $n$-th SES is occupied and
0 if it is empty. We use here Greek letters for the labelling of the
MESs and Latin letters for the labelling of the SESs. According to the
selection of the relevant SESs around the bias window they
will be of the form
\begin{equation}
      |{\bf \mu }\rangle =| \underbrace{1,1,....1}_{N_0 \,{\rm states}},
      i^{\mu}_{N_0+1},....,i^{\mu}_{N_{\rm max}},0,0,.....\rangle ,
\end{equation}
where $N_0$ denotes the highest SES below the bias window, but $N_\mathrm{max}$
is the number of the highest SES inside it.

The many-electron statistical operator of the whole system (leads and quantum wire)
obeys the  Liouville-von Neumann equation
\begin{equation}
      i\dot W(t)=[H(t),W(t)],\quad W(t<t_0)=\rho_\mathrm{L}\rho_\mathrm{R}\rho_\mathrm{S},
\label{L-vN}
\end{equation}
where the equilibrium density operator of the disconnected lead $l$ with
chemical potential $\mu_l$ is
\begin{equation}
      \rho_l=\frac{e^{-\beta (H_l-\mu_l N_l)}}{{\rm Tr}_l \{e^{-\beta(H_l-\mu_l N_l)}\}}.
\label{rho_l}
\end{equation}
$\rho_\mathrm{S}$ is the density operator of the isolated quantum wire ($t<t_0$) and serves as an
initial condition for the reduced density operator (RDO) defined as the (partial) trace over
the Fock space of the leads
\begin{equation}
      \rho(t)={\rm Tr}_\mathrm{L} {\rm Tr}_\mathrm{R} W(t),\quad \rho(t_0)=\rho_\mathrm{S}.
\label{rdo}
\end{equation}
We do not impose equilibrium condition on the leads after they have been coupled to the
system, the finite quantum wire, at $t>t_0$. This is different from other approaches
where one imposes $W(t)=\rho_L\rho_R\rho(t)$, (see for example Ref.\ \cite{Li05:205304}).
In the second order in the coupling Hamiltonian
the time evolution of the RDO is governed by the operator integro-differential equation 
\cite{Moldoveanu2008:GME}
\begin{equation}
      {\dot\rho}(t)=-\frac{i}{\hbar}[H_\mathrm{S},\rho(t)]
      -\frac{1}{\hbar^2}\sum_{l=\mathrm{L,R}}\int dq\:\chi^l(t)
      ([{\cal T}^l,\Omega_{ql}(t)]+h.c.),
      \label{GME}
\end{equation}
where we have introduced two operators to compactify the notation
\begin{eqnarray}
      &&\Omega_{ql}(t)=e^{-itH_\mathrm{S}} \int_{t_0}^tds\:\chi^l(s)
      \Pi_{ql}(s)e^{i(s-t)\varepsilon^l(q)}e^{itH_\mathrm{S}},\\
      &&\Pi_{ql}(s)=e^{isH_\mathrm{S}}\left ({\cal T}^{l{\dagger}}
      \rho(s)(1-f^l)-\rho(s){\cal T}^{l{\dagger}}f^l\right )e^{-isH_\mathrm{S}},
\label{PIql}
\end{eqnarray}
and a scattering operator ${\cal T}$ acting in the Fock space of the system
\begin{equation}
      {\cal T}^l(q)=\sum_{\alpha,\beta}{\cal T}_{\alpha\beta}^l(q)|{\bf \alpha}\rangle\langle {\bf \beta}| ,\quad
      {\cal T}_{\alpha\beta}^l(q)=\sum_aT^l_{aq}\langle {\bf \alpha}|d_a^{\dagger}|{\bf \beta}\rangle.
\label{Toperator}
\end{equation}
${\cal T}_{\alpha\beta}^l(q)$ describes the `absorption' of
electrons from the leads to the system and changes the
many-electron state of the latter from $\beta\rightarrow\alpha$.
The Fermi function of the SES labelled by $q\leftrightarrow (n_y^lq)$ in lead
$l$ is noted by $f^l$.
In order to derive Eq.\ (\ref{GME}) we have used the projection operator
${\cal P}=\rho_R(t_0)\rho_L(t_0){\mathrm Tr_RTr_L}$ to project out an
equation for the evolution of the system under the influence of the leads.
Though we do not do it here, we could equally well project out an 
evolution equation for the leads. This equation would tell
us how the electronic system in the leads evolves for $t>t_0$ without
assuming the central system to stay in equilibrium. Thus the Fermi
distribution present in Eq.\ (\ref{PIql}) is valid only for $t\leq t_0$
and the effective distribution for later times can only be found
from the RDO for the leads. This ansatz is not expected to change the
steady-state currents, but it will influence the transient regime.

Projecting (\ref{GME}) on the many-electron states
constructed from the relevant single-electron states in the extended bias-window results
in ${\cal N}=2^{N_{\rm max}-N_0}$ coupled integro-differential equations for the
matrix elements $\langle\alpha |\rho (t)|\beta\rangle$ of the reduced density
operator describing the time evolution of the open system, i.e.\ the finite
quantum wire coupled to the leads.

With the RDO it is now possible to calculate the statistical average of the
charge operator $Q_\mathrm{S}=e\sum_a d_a^{\dagger}d_a$ in the coupled quantum wire
\begin{eqnarray}
      \langle Q_\mathrm{S}(t)\rangle&=&{\rm Tr}\{W(t)Q_\mathrm{S}\}
      ={\rm Tr}_\mathrm{S}\{ [{\rm Tr}_\mathrm{LR}W(t)]Q_\mathrm{S} \}\nonumber\\
      &=&{\rm Tr}_\mathrm{S}\{\rho(t)Q_\mathrm{S} \}
      =e\sum_{a,\mu} i^{\mu}_a \, \langle\mu | \rho(t) | \mu\rangle ,
\end{eqnarray}
with the traces assumed over the Fock space.
We are also interested in the average spatial distribution of the time-dependent charge
\begin{equation}
      \langle Q_\mathrm{S}({\bf r},t)\rangle = e\sum_{ab}\sum_{\mu\nu}
      \Psi^*_a({\bf r})\Psi_b({\bf r})\rho_{\mu\nu}(t)\langle\nu |d^\dagger_ad_b|\mu\rangle .
\label{Qxy}
\end{equation}
The {\it net} current flowing into the sample is
\begin{equation}
      \Delta\langle J_\mathrm{T}(t)\rangle=\langle J_\mathrm{T}^\mathrm{L}(t)\rangle
      -\langle J_\mathrm{T}^\mathrm{R}(t)\rangle =\frac{d\langle Q_\mathrm{S}(t)\rangle}{dt}
      =e\sum_a \sum_{\mu} i^{\mu}_a \, \langle\mu | \dot\rho(t) | \mu\rangle \,.
\end{equation}
Through the GME (\ref{GME}) it is possible to identify the contribution
of each SES in the system to the current from the left lead or into the
right lead \cite{Moldoveanu2008:GME}.
More precisely, Eq.\ (\ref{GME}) gives us an expression for $\dot\rho(t)$. 
The trace of the first part, the commutator of $\rho$ and $H_\mathrm{S}$ vanishes.
The summation over leads in the second part allows us to identify the contribution
to or from each lead.

The iterative scheme to solve the GME has been described in an earlier
publication \cite{Moldoveanu2008:GME}. Due to the commutator structure of
the GME (\ref{GME}) the conservation of probabilities can be verified,
i.e.\ ${\rm Tr}_\mathrm{S} {\dot\rho}(t)=0$. At any time step in the numerical
iteration we check the conservation of probability.

%%
% Results
%--------
\section{Results and discussion}
An advantage of the GME formalism is the freedom to select the initial
state of the system in any way compatible with the description via the
RDO. We can thus choose the system to be initially empty, or occupied by
any number of electrons influenced by the equilibrium state in either
lead. More complex nonequilibrium initial states are also possible.

In our calculation we assume GaAs parameters, $m^*=0.067m_e$, $\kappa = 12.4$,
and set $t_0=0$.
The characteristic energy of the parabolic confinement for both the finite
quantum wire and the leads is assumed $\hbar\Omega_0=1.0$ meV.
The coupling between the leads and the quantum wire is described by the function
\begin{equation}
      \chi^{\mathrm{L,R}}(t)=\left\{ \begin{array}{cl}
      \left(1-\frac{2}{e^{\alpha^{\mathrm{L,R}} t}+1}\right) &\mbox{if $t>0$} \\
       0 &\mbox{if $t\leq 0$}
       \end{array} \right.
\label{chi}
\end{equation}
with $\alpha^l = 1.0$ ps$^{-1}$. We fix the temperature of the reserviors at
$T=0.5$ K. The parameters determining the
coupling of the subsystems in the function $g^\mathrm{L,R}$ (\ref{gl}) are $\delta_1a_w^2=1.0$,
$\delta_2^la_w^2=2.0$, and $\Delta_E^l=1.0$ meV in order to let the coupling strength be
determined by the behavior of the eigenstates of the quantum wire and the leads only
very close to the contact region at $x=\pm L_x/2$.
For the numerical coupling constant $g_0^l$ in (\ref{gl}) we select the
value $g_0^la_w^{3/2}=g_{00}^la_w^{3/2}$ with $g_{00}^l=10.0$ meV for both leads, or a different
value stated for a particular calculation. (The rather unusual dimension of the
numerical coupling constant comes from the fact that the $x$-part of the wave function
in the leads is only $\delta$-normalizable, see Appendix A). The numerical value of
$g_{00}^l$ has per se a limited significance, but below we shall explicitly show the
effective coupling of the states in the leads and the system graphically.

A time step $\Delta t=0.01519$ ps
was selected in order to allow for time integration to relatively large times without
any visible loss of accuracy on the scales presented in the figures to follow.
As will be shown in subsequent figures we include part of 3-4 energy subbands in the
$q$-integration in the GME (\ref{GME}). For the wavevector cut-off $qa_w=4.0$
it has proven essential to use a large enough number of $qa_w$-integration
points to guarantee high accuracy for large times.
Within each subband we use a
4-point Gauss integration repeated 180 times. This requirement weighs heaviest
in increasing the needed CPU-time.

Embedded in the finite quantum wire we have a quantum dot or an antidot represented by the
Gaussian potential
\begin{equation}
      V(\mathbf{r})=V_0\exp{\left\{-[\beta_x (x-x_0)]^2-[\beta_y(y-y_0)]^2\right\}},
\label{V}
\end{equation}
with $\beta_x=\beta_y=0.03$ nm$^{-1}$, and $V_0$ and the spatial off-sets to be stated
later.

\subsection{Pure finite wire}
Before presenting results for the quantum wire with an embedded subsystem we use
results for the time-dependent transport through the pure finite quantum wire to
familiarize us with the system.

\begin{figure}[htbq]
      \begin{center}
      \includegraphics[width=0.54\textwidth,angle=0]{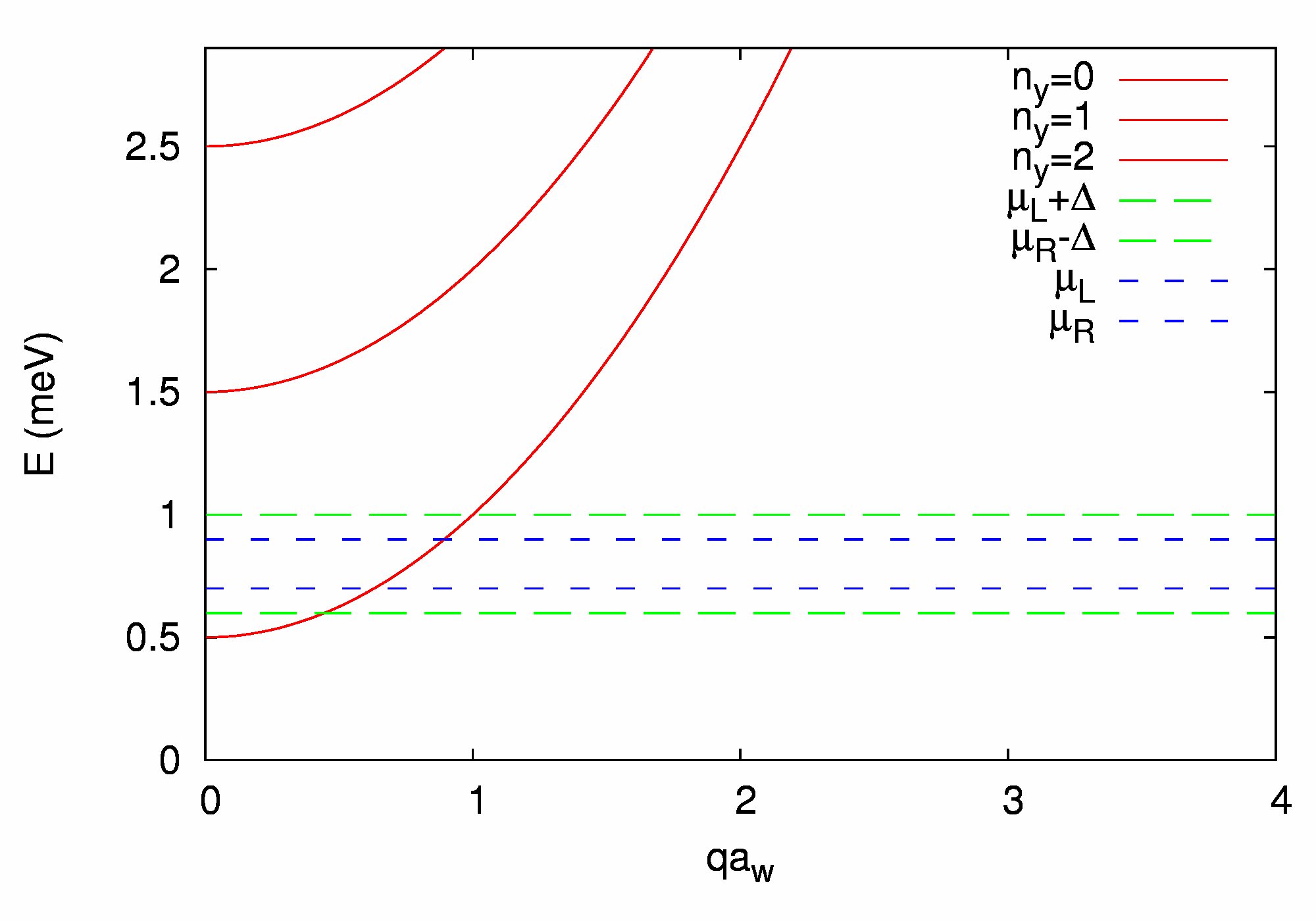}
      \end{center}
      \caption{The energy spectrum of the leads (solid red) vs.\
               the scaled wavevector $qa_w$, and
               the chemical potential in each lead $\mu_L=0.9$ meV, $\mu_R=0.7$ meV,
               and the window of relevant states [$\mu_R-\Delta$, $\mu_L+\Delta$]
               for $\Delta =0.1$ meV, $\hbar\Omega_0=1.0$ meV.}
      \label{E_leads}
\end{figure}

\begin{figure}[htbq]
      \begin{center}
      \includegraphics[width=0.54\textwidth,angle=0]{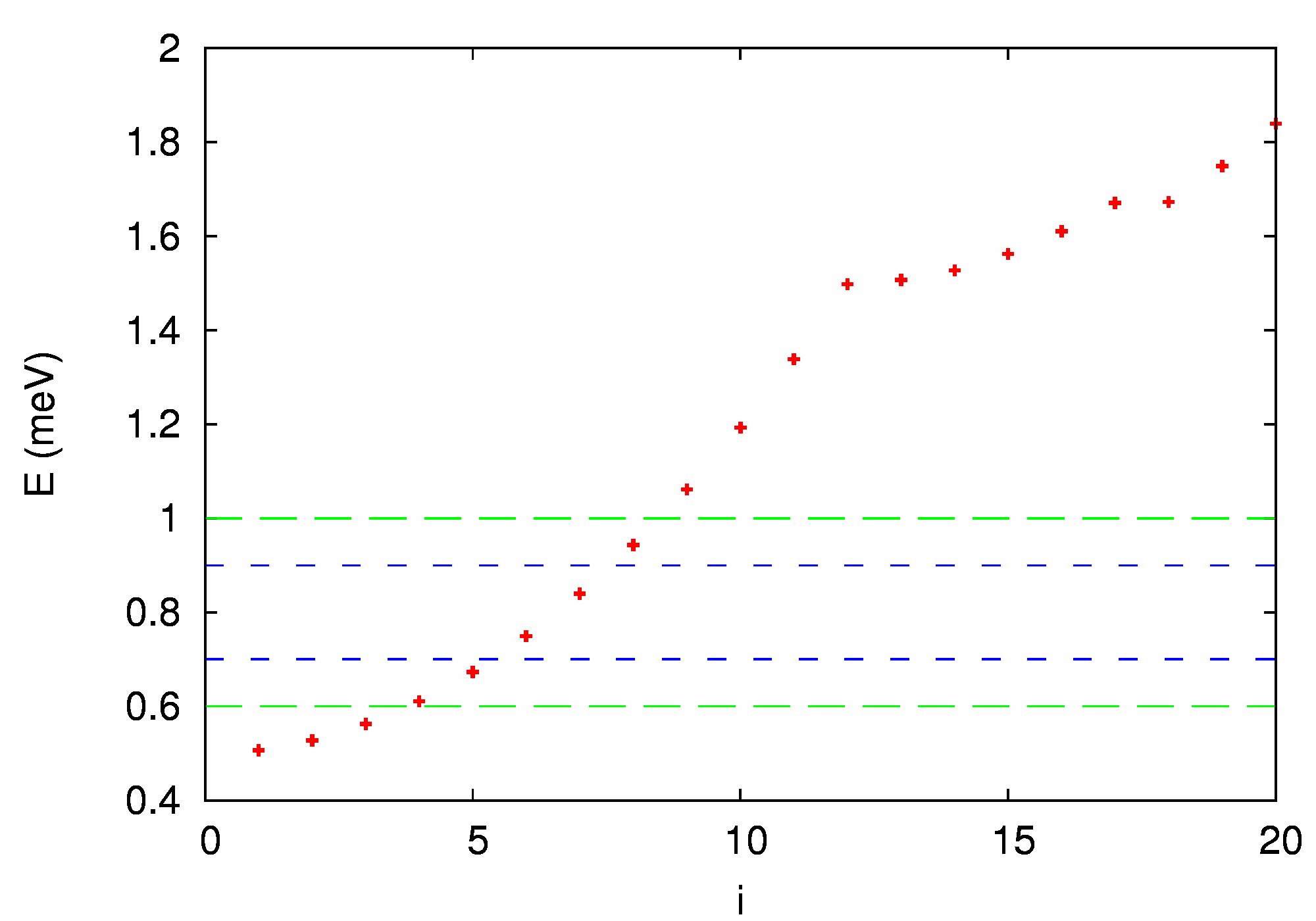}
      \end{center}
      \caption{The energy spectrum of the system (red dots)
               vs.\ the SES state number $i$, and
               the chemical potential in each lead $\mu_L=0.9$ meV, $\mu_R=0.7$ meV,
               and the window of relevant states [$\mu_R-\Delta$, $\mu_L+\Delta$]
               for $\Delta =0.1$ meV. $L_x=900$ nm, $\hbar\Omega_0=1.0$ meV.}
      \label{E_HS}
\end{figure}

The energy spectrum of the
leads is shown in Fig.\ \ref{E_leads} together with the chemical potential in
each lead, $\mu_L=0.9$ meV, $\mu_R=0.7$ meV, and the limits $\mu_R-\Delta$ and
$\mu_L+\Delta$ with $\Delta =0.1$ meV defining the window of relevant states
around the applied bias $eV_\mathrm{bias}=\mu_L-\mu_R=0.2$ meV.
The maximum energy for each subband shown in the graph indicates the corresponding
maximum wavevector in the $qa_w$-integration of the GME (\ref{GME}). The energy spectrum
of the $900$ nm long quantum wire is shown in Fig.\ \ref{E_HS}.
together with the extended bias window [$\mu_R-\Delta$, $\mu_L+\Delta$] containing 5 SESs,
the relevant SESs for including in the transport calculation and construction of the MESs.
The actual bias window contains 2 SESs. The finite quantum wire is long
enough compared to its effective width to show a clear indication of formation of energy
subbands in the energy range shown in the figure.

\begin{figure}[htbq]
      \begin{center}
      \includegraphics[width=0.49\textwidth,angle=0]{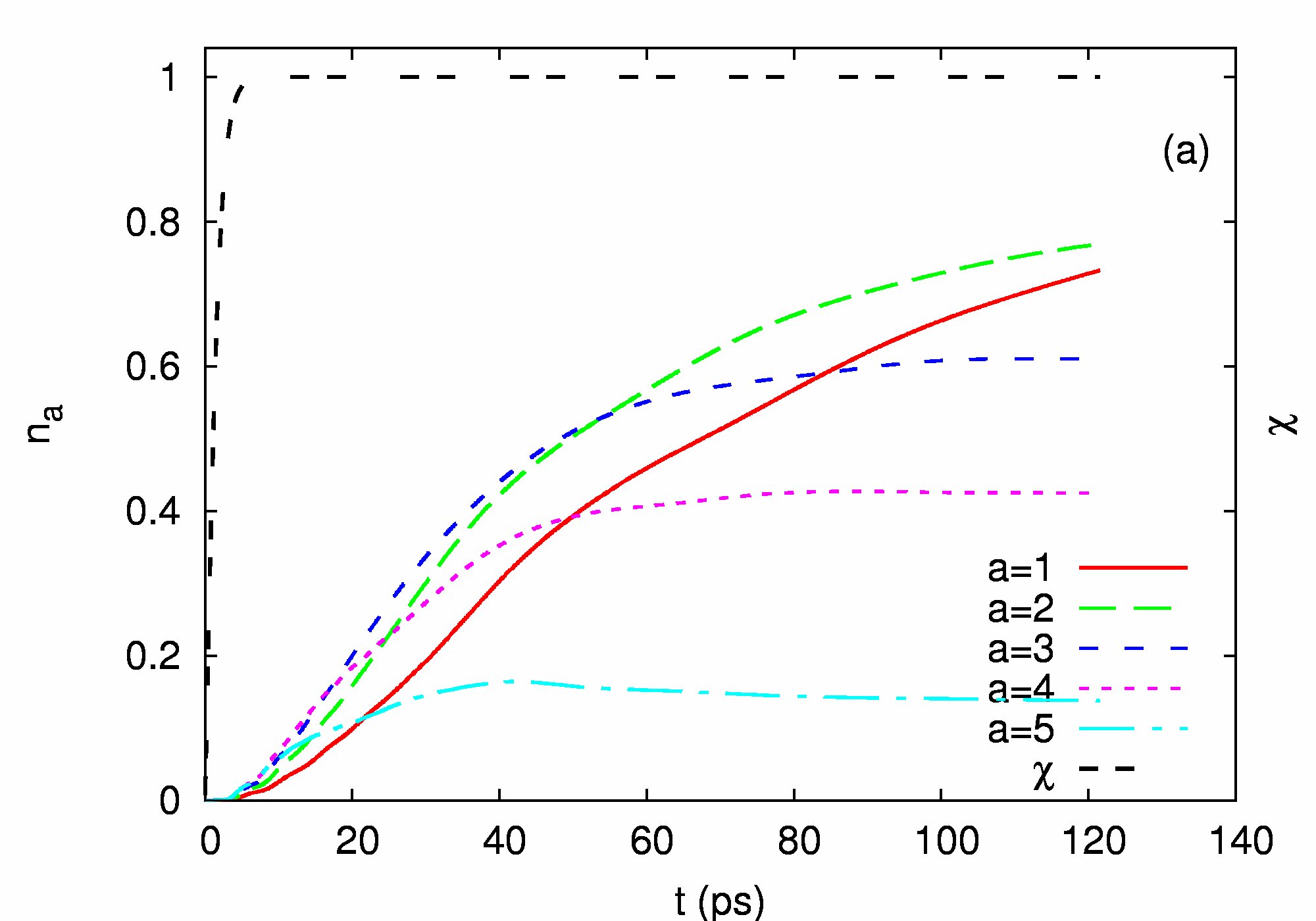}
      \includegraphics[width=0.49\textwidth,angle=0]{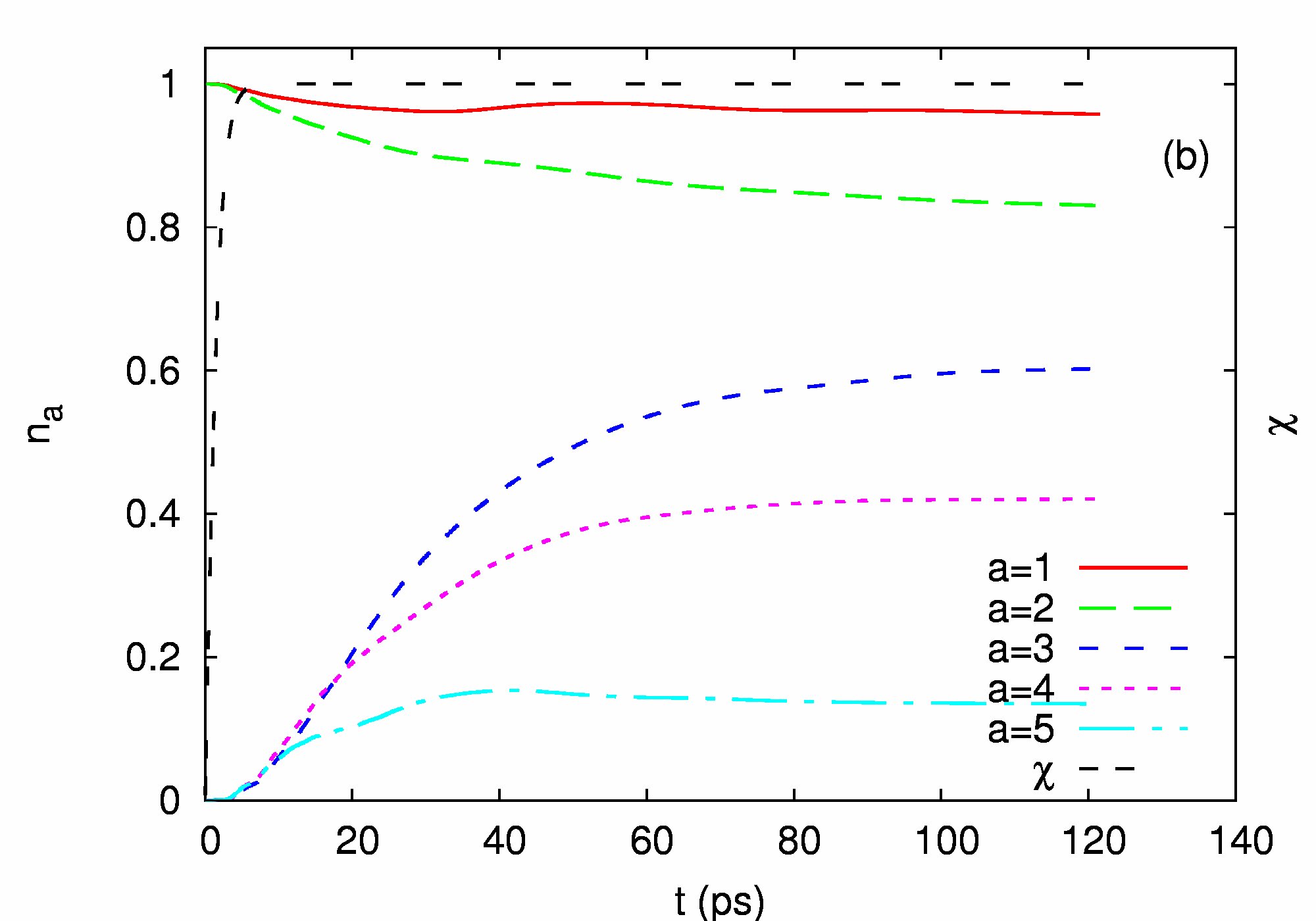}
      \end{center}
      \caption{The time dependent occupation of the relevant
               SESs for an initially empty system $\mu_0=1$ (a), and for
               a system occupied with initially 2 electrons ($\mu_0=4$) in equilibrium (b).
               The time coupling function $\chi^l$ is shown for reference.
               $L_x=900$ nm, $\hbar\Omega_0=1.0$ meV, $g_{00}^l=10.0$ meV,
               $\Delta_E^l=1.0$ meV, and $T=0.5$ K.}
      \label{n_a}
\end{figure}

The time-dependent occupation of the relevant SESs is shown in Fig.\ \ref{n_a} in
comparison with the time-dependent coupling function $\chi^\mathrm{L,R}$.

The numbering of the relevant SESs with the index $a$ is in the order of increasing energy.
In the initially empty system the many-electron state is labelled by $\mu_0=1$,
and the two-electron state with the electrons in the two lowest SES is labelled
by $\mu_0=4$ \cite{Moldoveanu2008:GME}.
We note that some of the higher SES seem to reach a steady state fractional
filling fast, while the ones lower in energy are still increasing their occupation
at times as large as 120 ps. Two effects contribute to this, the higher lying states
are stronger coupled to states in the leads as we will show below, and they can
conduct faster. The time-dependent GME formalism introduces energy dissipation from
the finite quantum wire into the leads. At the finite temperature,
$T=0.5$ K, we see that in case of two electron initially in the system there is a finite
but small probability for the electrons to get out of the system, even though they
have energy below the actual bias window. As expected, the system looses the electron
occupying the SES closer to the bias window with a higher probability.

\begin{figure}[htbq]
      \begin{center}
      \includegraphics[width=0.54\textwidth,angle=0]{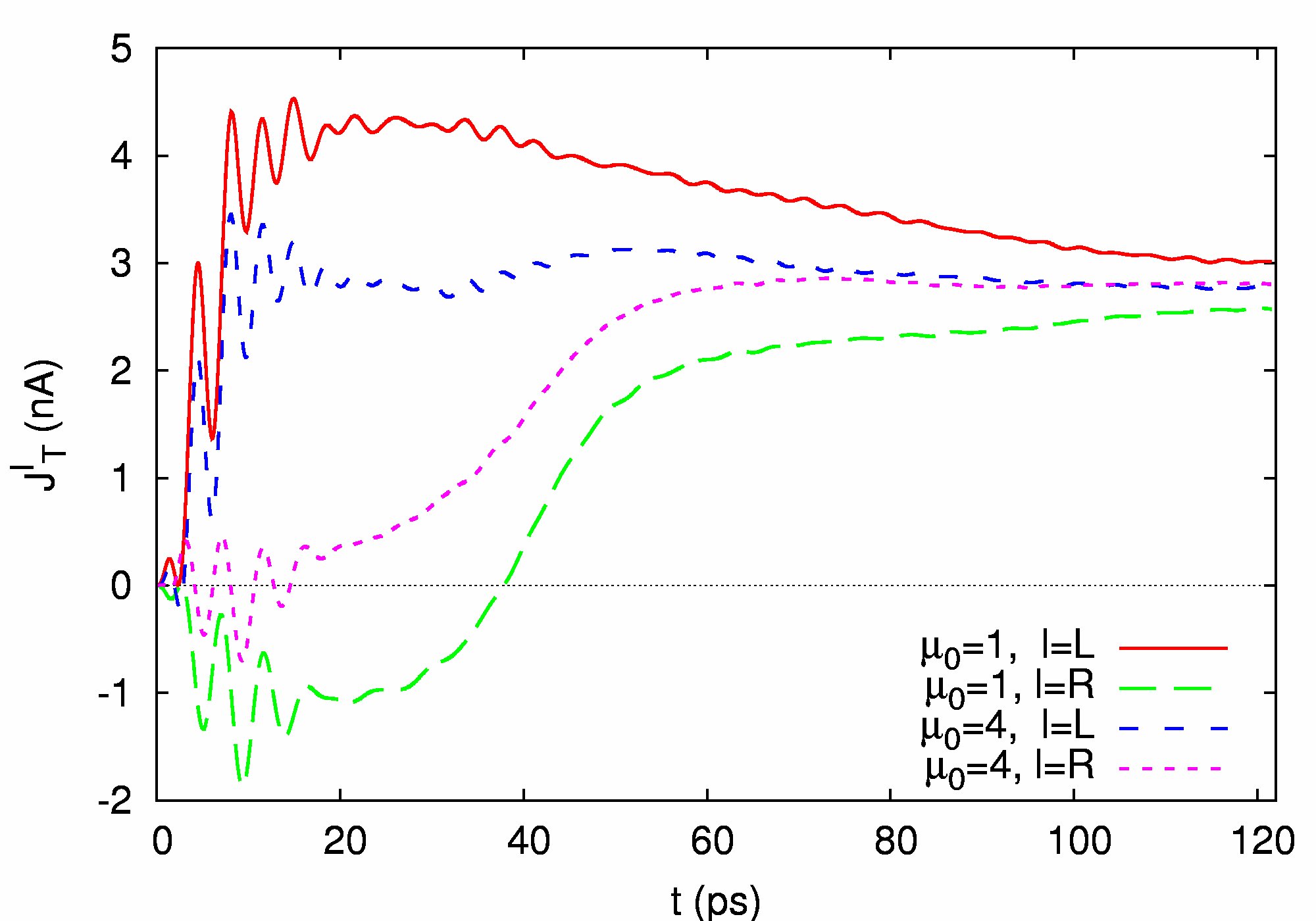}
      \end{center}
      \caption{The total current entering the system from
               the left lead $J_T^L(t)$, and the total current exiting the system into
               the right lead $J_T^R(t)$ vs.\ time for an initially empty system
               $\mu_0=1$ and a system with two electrons initially in equilibrium
               $\mu_0=4$.
               $L_x=900$ nm, $\hbar\Omega_0=1.0$ meV, $g_{00}^l=10.0$ meV,
               $\Delta_E^l=1.0$ meV, and $T=0.5$ K.
               A positive sign of the current indicates a flow of electrons
               with charge $e$ from left to right.}
      \label{J_T}
\end{figure}

The time-dependent total current injected into the system from the left lead $J_T^L(t)$ and the
total current leaving the system into the right lead $J_T^R(t)$ are shown in Fig.\ \ref{J_T}.
Here we see that in case of the initially empty system, $\mu_0=1$, the current
in the right lead, $J_T^\mathrm{R}$, is negative meaning that
it is directed into the system for $t<40$ ps.
The system is supplied with electrons from both ends initially before it reaches
a steady state with constant current through it. In case of the system initially occupied by
two electrons, $\mu_0=4$, we do not have a net transfer of charge from the right lead initially,
but we see fluctuations in $J_T^\mathrm{R}$ for $t<20$ ps before it turns positive
when a net current is flowing through the system.
Not surprisingly, the current has a maximum value when the occupation is
changing the fastest for the system, see Fig.\ \ref{n_a}.

\begin{figure}[htbq]
      \begin{center}
      \includegraphics[width=0.50\textwidth,angle=0,viewport=10 22 350 250,clip]{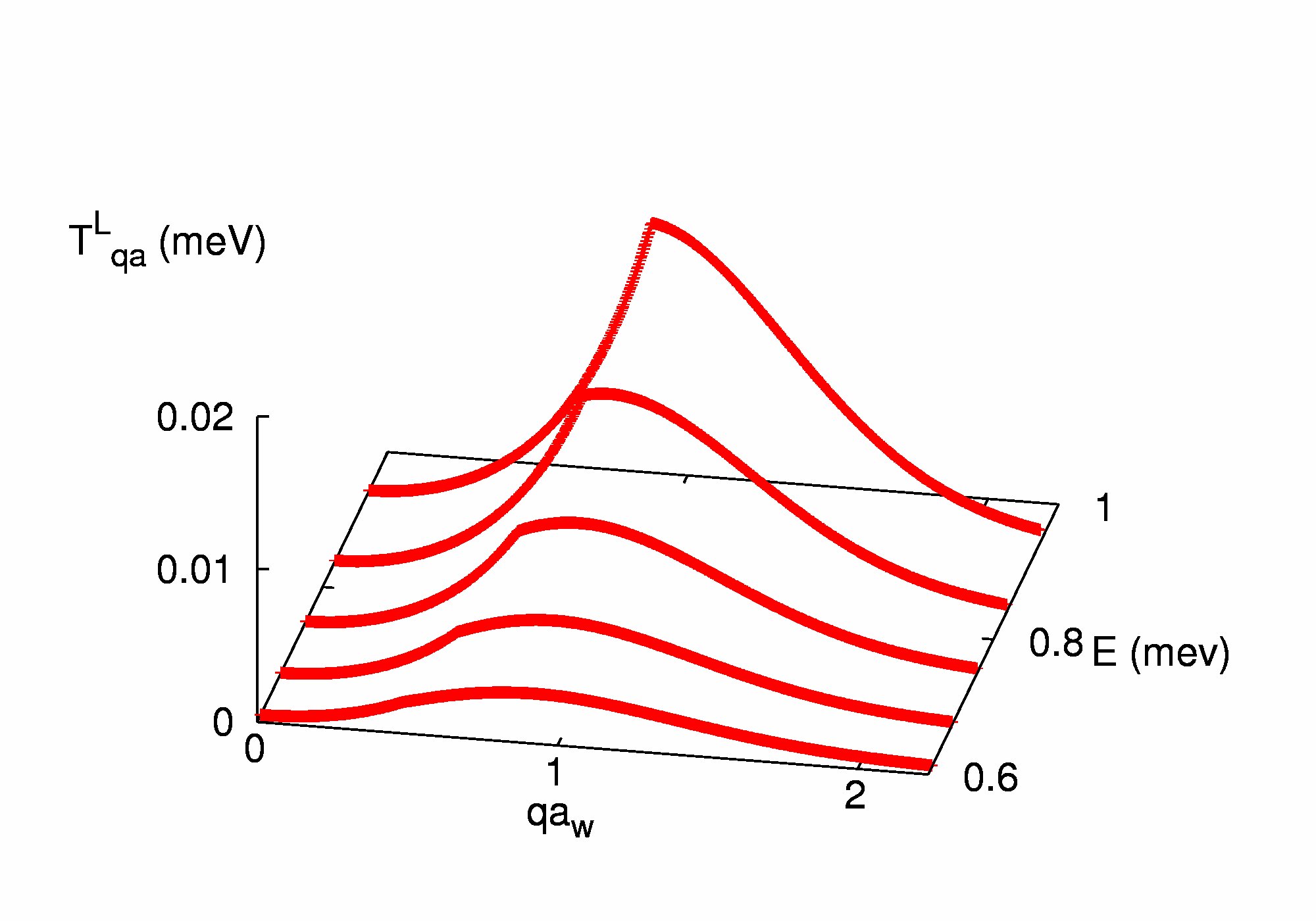}\\
      \includegraphics[width=0.49\textwidth,angle=0,viewport=10 20 350 140,clip]{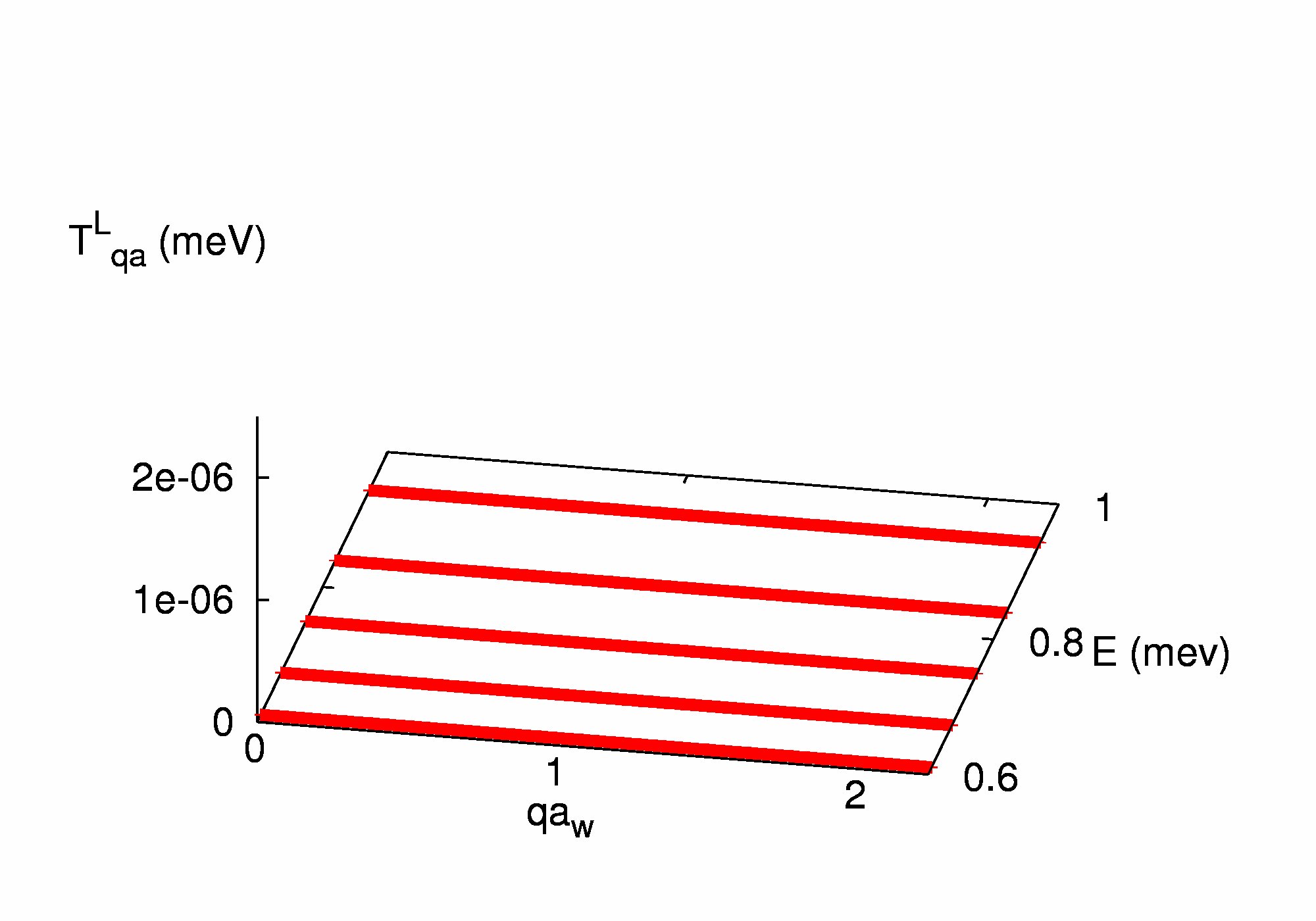}
      \includegraphics[width=0.49\textwidth,angle=0,viewport=10 20 350 176,clip]{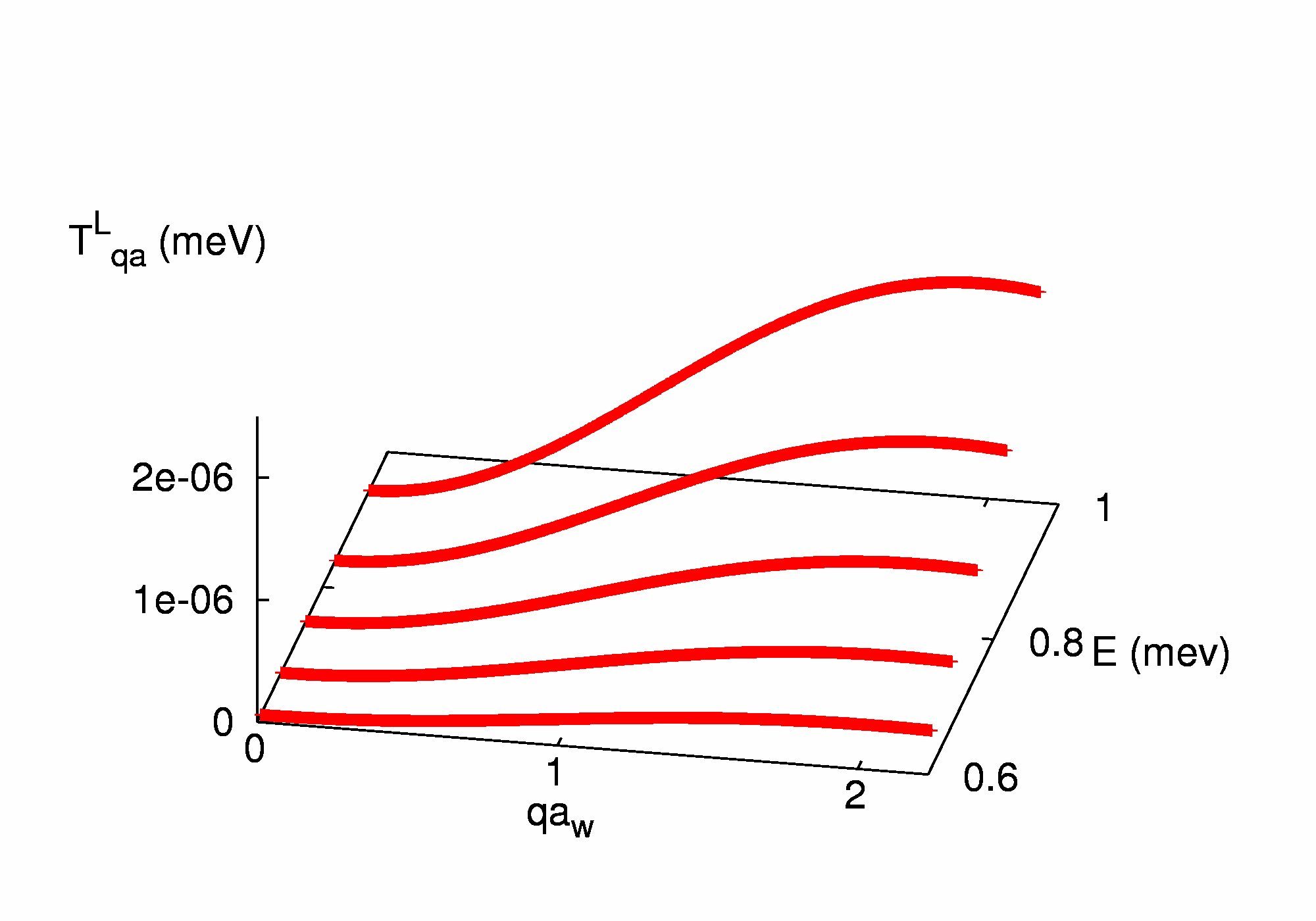}
      \end{center}
      \caption{The coupling between a state $a$ in the relevant
               extended bias window and state $qn_y$ in the left (and the right) lead
               for $n_y=0$ (top panel), $n_y=1$ (lower left panel), and  $n_y=2$ (lower right).
               $L_x=900$ nm, $\hbar\Omega_0=1.0$ meV, $g_{00}^l=10.0$ meV,
               $\Delta_E^l=1.0$ meV.}
      \label{T_L}
\end{figure}

The coupling strength tensor introduced earlier (\ref{T_aq}) gives the coupling
between a state $a$ in the relevant extended bias window and a state $qn_y$ in the
leads (here the coupling is the same for left and right lead). In Fig.\ \ref{T_L}
we see the tensor for the 3 lowest subbands of the leads labelled by $n_y$.
The maxima correspond to resonant tunneling when $E_a=\epsilon_{n_y}(q)$. The type of
coupling selected (\ref{T_aq}) reproduces an effect well known in models for
multimode transport in quasi-one dimensional quantum wires built on the Lippmann-Schwinger
scattering approach \cite{Bardarson04:245308}, i.e.\ the parity of the subband wavefunctions
in case of a symmetric system forbids coupling of the nearest neighboring subbands. Therefore, the
coupling of the relevant states here all lying in the first subband of the finite quantum
wire to the second subband of the leads is vanishing, as is seen in the center subfigure
of Fig.\ \ref{T_L}. The coupling to the third subband is reduced by the exponential term
in energy of Eq.\ (\ref{T_aq}).

\begin{figure}[htbq]
      \begin{center}
      \includegraphics[width=0.49\textwidth,angle=0,viewport=20 15 360 210,clip]{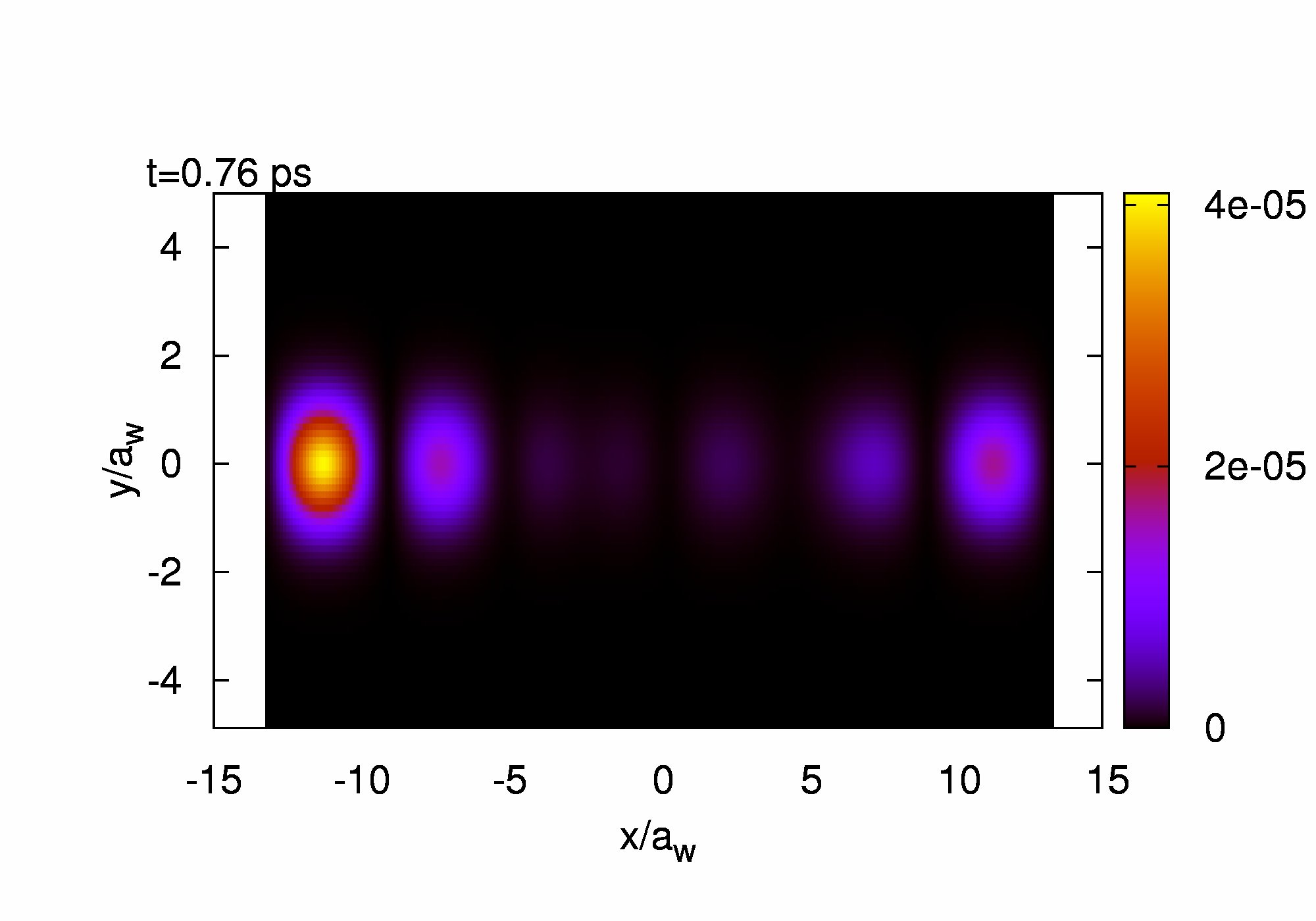}
      \includegraphics[width=0.49\textwidth,angle=0,viewport=20 15 360 210,clip]{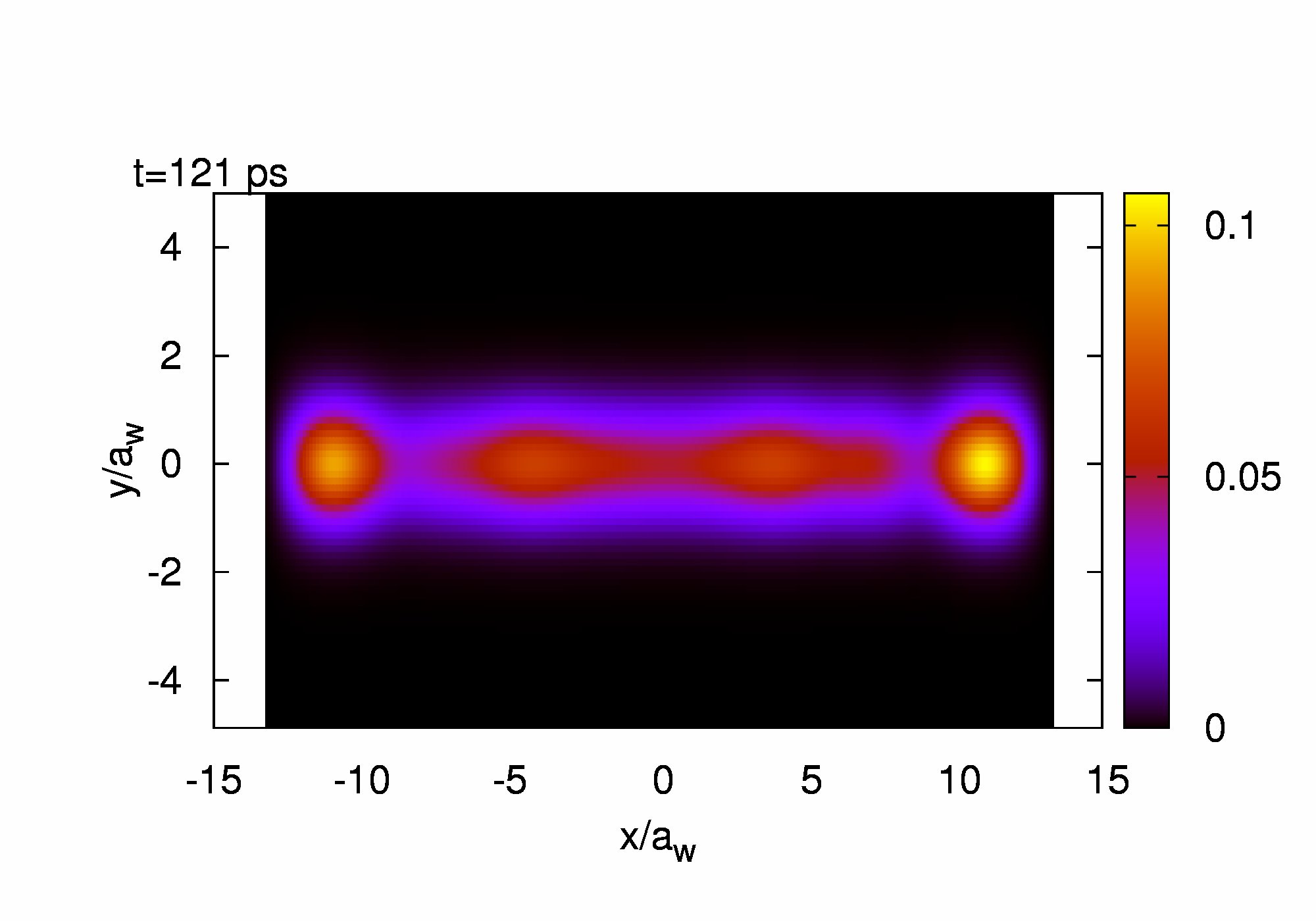}
      \end{center}
      \caption{The average spatial charge distribution for the MES constructed from the
               5 relevant SESs in the extended bias window for $t=0.76$ ps (left panel), and
               $t=121$ ps (right panel). Note the huge difference in scale.
               The system is initially empty $\mu_0=1$.
               $L_x=900$ nm, $\hbar\Omega_0=1.0$ meV, $g_{00}^l=10.0$ meV,
               $\Delta_E^l=1.0$ meV, and $T=0.5$ K.}
      \label{Qnn}
\end{figure}

The RDO can be used to calculate the average spatial charge distribution (\ref{Qxy})
of the MES in the finite quantum wire at any instant of time. In Fig.\ \ref{Qnn} we
show it soon after the coupling of the system to the leads, and again when the system
is close to reaching a steady state.
Just as we have seen in Fig.\ \ref{J_T} of the total current in the right and left
leads, initially the probability density increases in the empty finite wire from both sides with the
higher bias to the left supplying it faster there. The steady state attained in the end
is clearly a mixed state with contribution from all of the available SES, and the
coupling to the leads maintains a higher probability at the ends of the finite wire.

\subsection{Finite wire with an embedded subsystem}
\begin{figure}[htbq]
      \begin{center}
      \includegraphics[width=0.54\textwidth,angle=0]{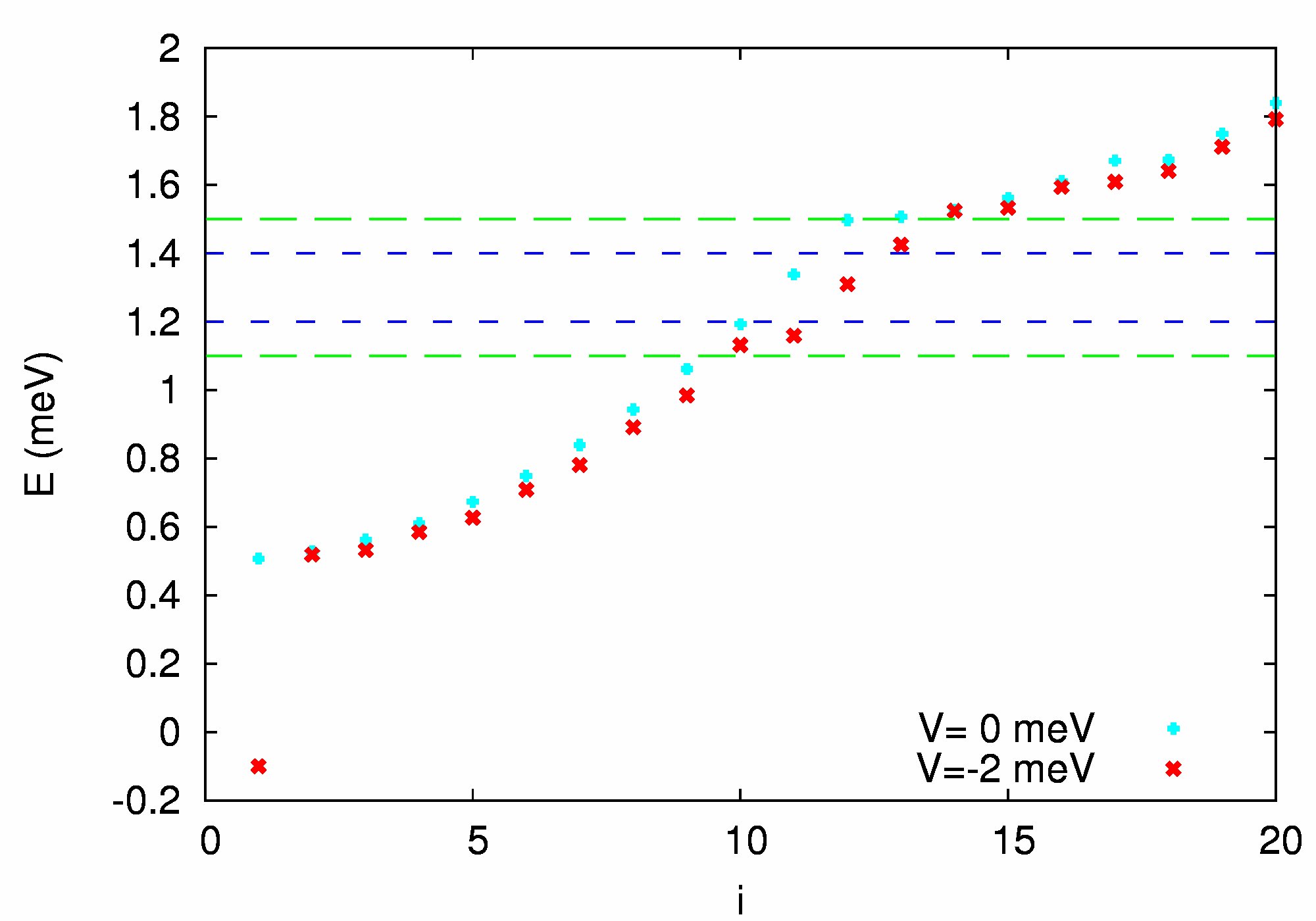}
      \end{center}
      \caption{The energy spectrum of the system (red $\times$)
               vs.\ the SES state number $i$ for the system with an embedded
               off-centered Gaussian well compared with the spectrum of a
               purely parabolic system (blue $+$), and
               the chemical potential in each lead $\mu_L=1.4$ meV, $\mu_R=1.2$ meV,
               and the window of relevant states [$\mu_R-\Delta$, $\mu_L+\Delta$]
               for $\Delta =0.1$ meV. $L_x=900$ nm, $\hbar\Omega_0=1.0$ meV, $V_0=-2.0$ meV,
               $\beta_x=\beta_y=0.03$ nm$^{-1}$, $y_0=30$ nm, and $x_0=0$.}
      \label{E_HS_V2m_y0p30}
\end{figure}

\begin{figure}[htbq]
      \begin{center}
      \includegraphics[width=0.49\textwidth,angle=0]{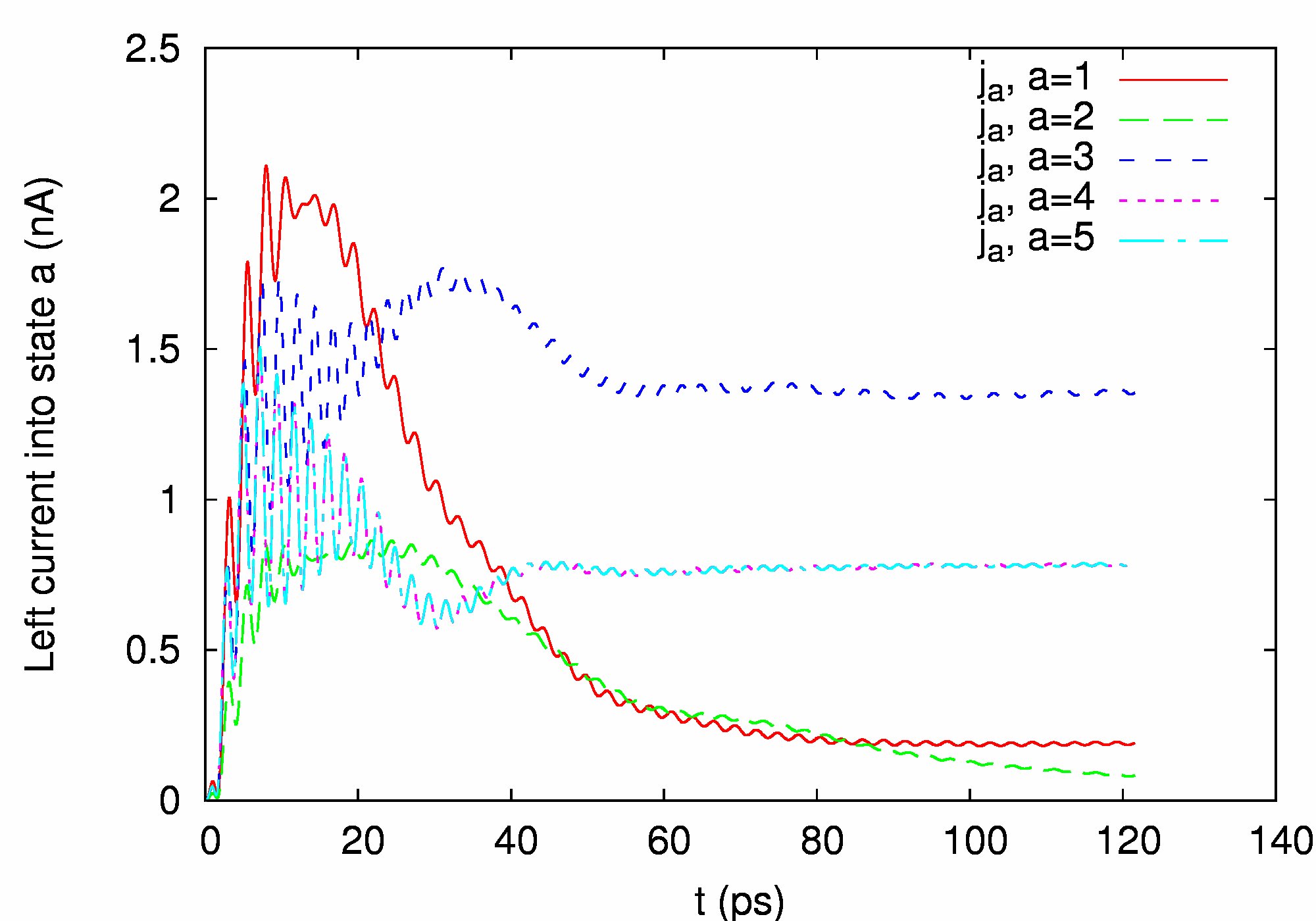}
      \includegraphics[width=0.49\textwidth,angle=0]{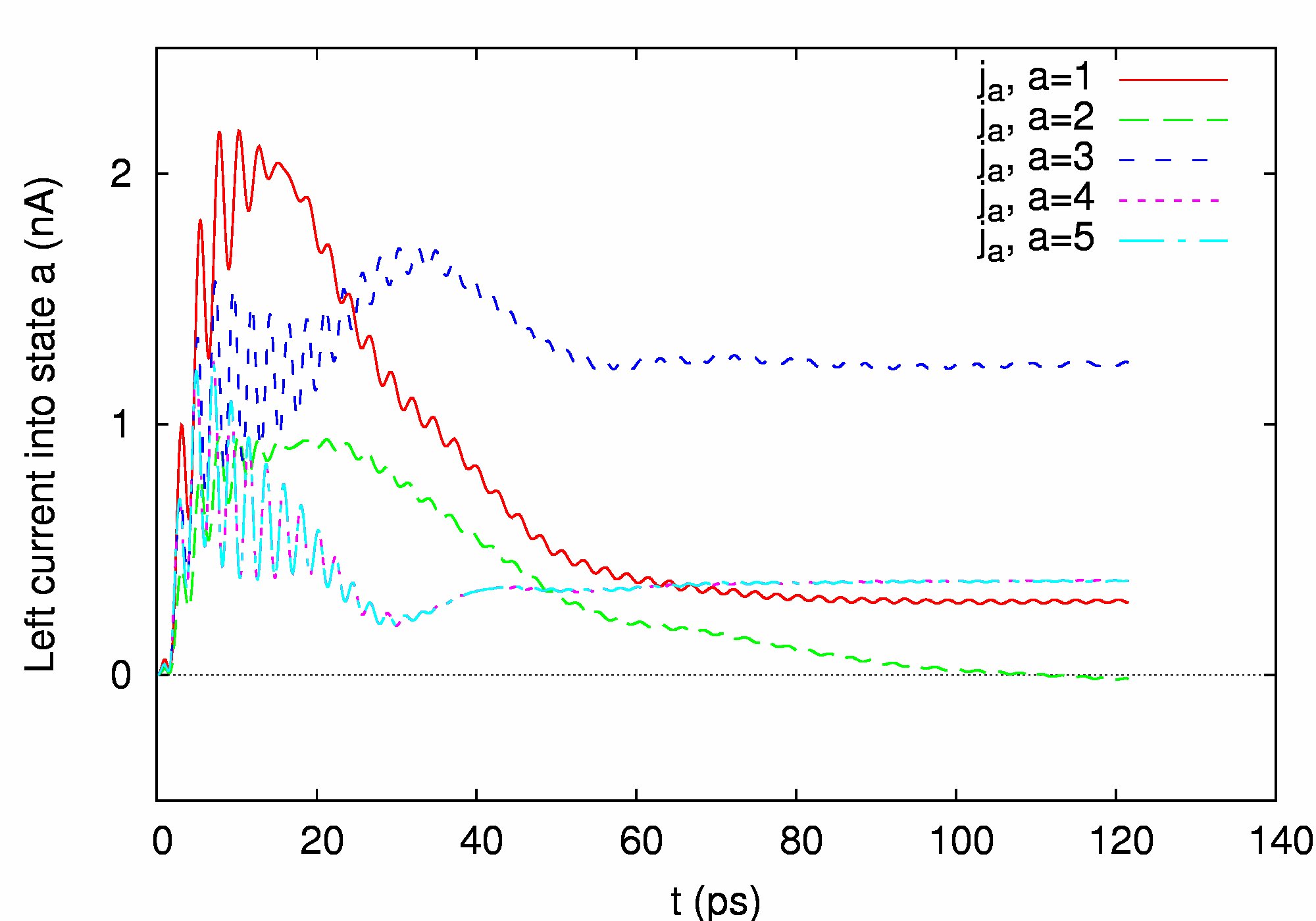}
      \end{center}
      \caption{The time-dependent current from the left lead into the SES
               $a$ for the system with an embedded off-centered Gaussian well
               at $y_0=30$ nm (left panel), and $y_0=60$ nm (right panel). $\Delta =0.15$ meV.
               $L_x=900$ nm, $\hbar\Omega_0=1.0$ meV, $V_0=-2.0$ meV,
               $\beta_x=\beta_y=0.03$ nm$^{-1}$, $g_{00}^l=7.5$ meV, and $x_0=0$.}
      \label{Lja_V2m_y0p3060_g15}
\end{figure}

\begin{figure}[htbq]
      \begin{center}
      \includegraphics[width=0.54\textwidth,angle=0]{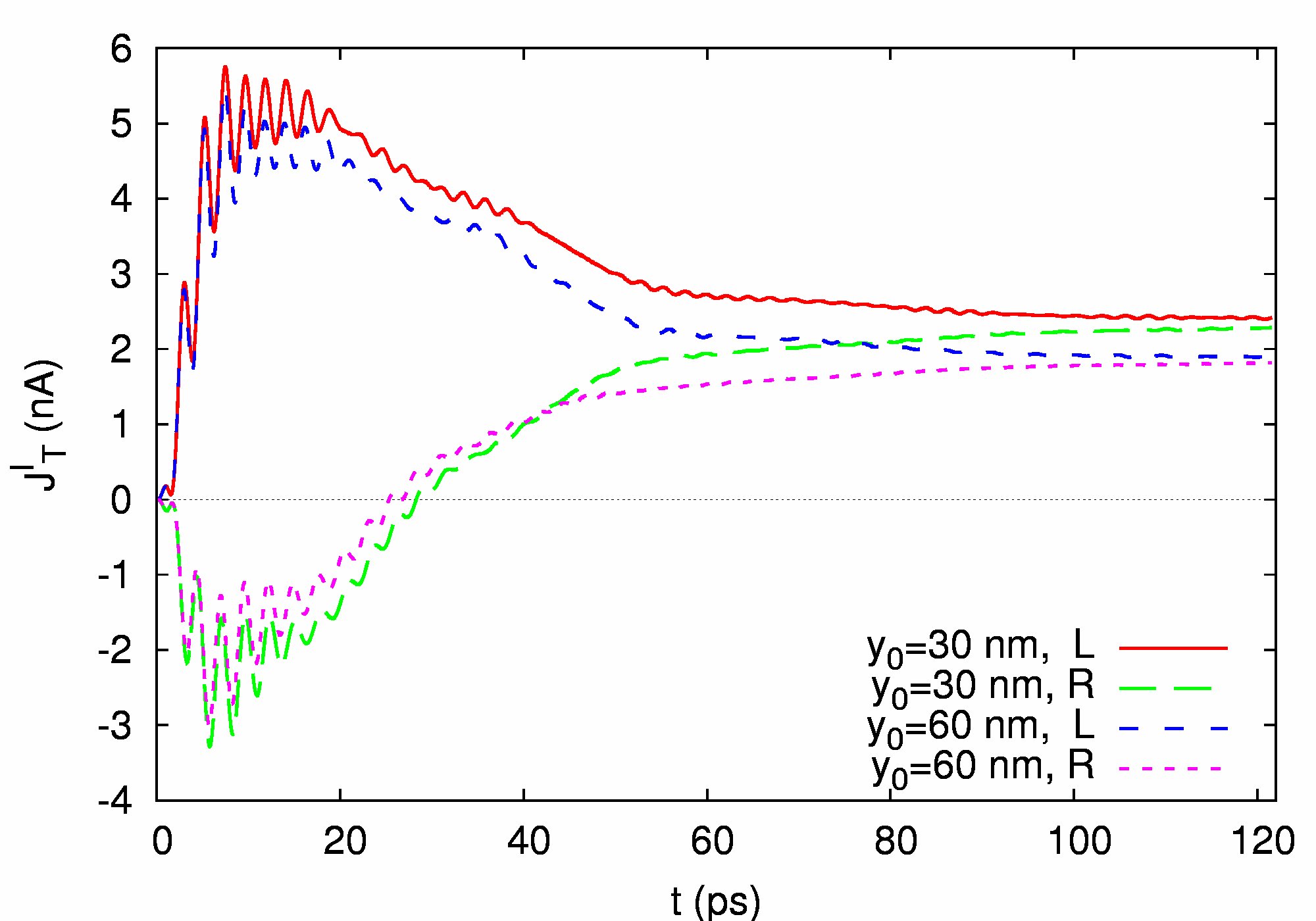}
      \end{center}
      \caption{The total left and right current vs.\ time
               for the system with an embedded off-centered Gaussian well.
               $\Delta =0.15$ meV.
               $L_x=900$ nm, $\hbar\Omega_0=1.0$ meV, $V_0=-2.0$ meV,
               $\beta_x=\beta_y=0.03$ nm$^{-1}$, $g_{00}^l=7.5$ meV, and $x_0=0$.
               A positive sign of the current indicates a flow of electrons
               with charge $e$ from left to right.}
      \label{JT_V2m_y0p3060_g15}
\end{figure}

We now continue our exploration of the effects of the geometry of the system and
leads on the description of the time-dependent transport by the GME formalism by
introducing a Gaussian potential into the finite quantum wire. In Fig.\ \ref{E_HS_V2m_y0p30}
the energy spectrum of the SES is shown for an off-centered Gauss well with
parameters $V_0=-2.0$ meV, and $\beta_x=\beta_y=0.03$ nm$^{-1}$ for the potential
given by Eq.\ (\ref{V}). For all the cases of an off-set Gaussian potential we
are only considering an off-set in the $y$-direction and we shall keep $x_0=0$.
In the figure we indicate the extended bias-window for the relevant states
by $\Delta =0.1$ meV, but in the calculations to
follow we will often use a larger value for $\Delta$ to be stated in each case.
We start selecting the location of the bias window just below and touching
the second subband of the system. Below we shall discuss the character of
the SESs, but first we investigate the current through the SESs of the
system. Experimentally only the total current entering or leaving the system
is measureable, but the current through each SES can give us an insight into
the transport processes in the system. The current from the left lead into
each relevant SES is displayed in Fig.\ \ref{Lja_V2m_y0p3060_g15} for two
different off-sets of the Gaussian well, the upper panel for the smaller
off-set, $y_0=30$ nm, and the lower panel for the larger one, $y_0=60$ nm.
We see that the current into the lowest state in energy, $a=1$, which is below
the chemical potential in both leads is highest the first 60 ps while
it is reaching its steady state value. After this period the
current reaches a steady state value that can be verified by checking that
the current leaving the system into the right lead (not shown here) has the same value.
The same can be said about the current through the states $a=3,4$, but the current
through state $a=2$ situated just below $\mu_R$ does not seem to reach a steady
state value in the time interval shown. Indeed, the current through $a=2$, $j_2$,
seems to be vanishing here. We shall investigate this further just below, but
first we show in Fig.\ \ref{JT_V2m_y0p3060_g15} how indeed, the total
current into the system from the left and the total current leaving the
system on the right reach the same values soon after $t=60$ ps.
Here is also clear that during the first $20$ ps the total current in the right
lead is directed into the system supplying it with electrons. Though we do not
show it here, the same is even true for all the partial right currents in this
time interval, though strongly decreasing for SESs with higher energy.

\begin{figure}[htbq]
      \begin{center}
      \includegraphics[width=0.49\textwidth,angle=0]{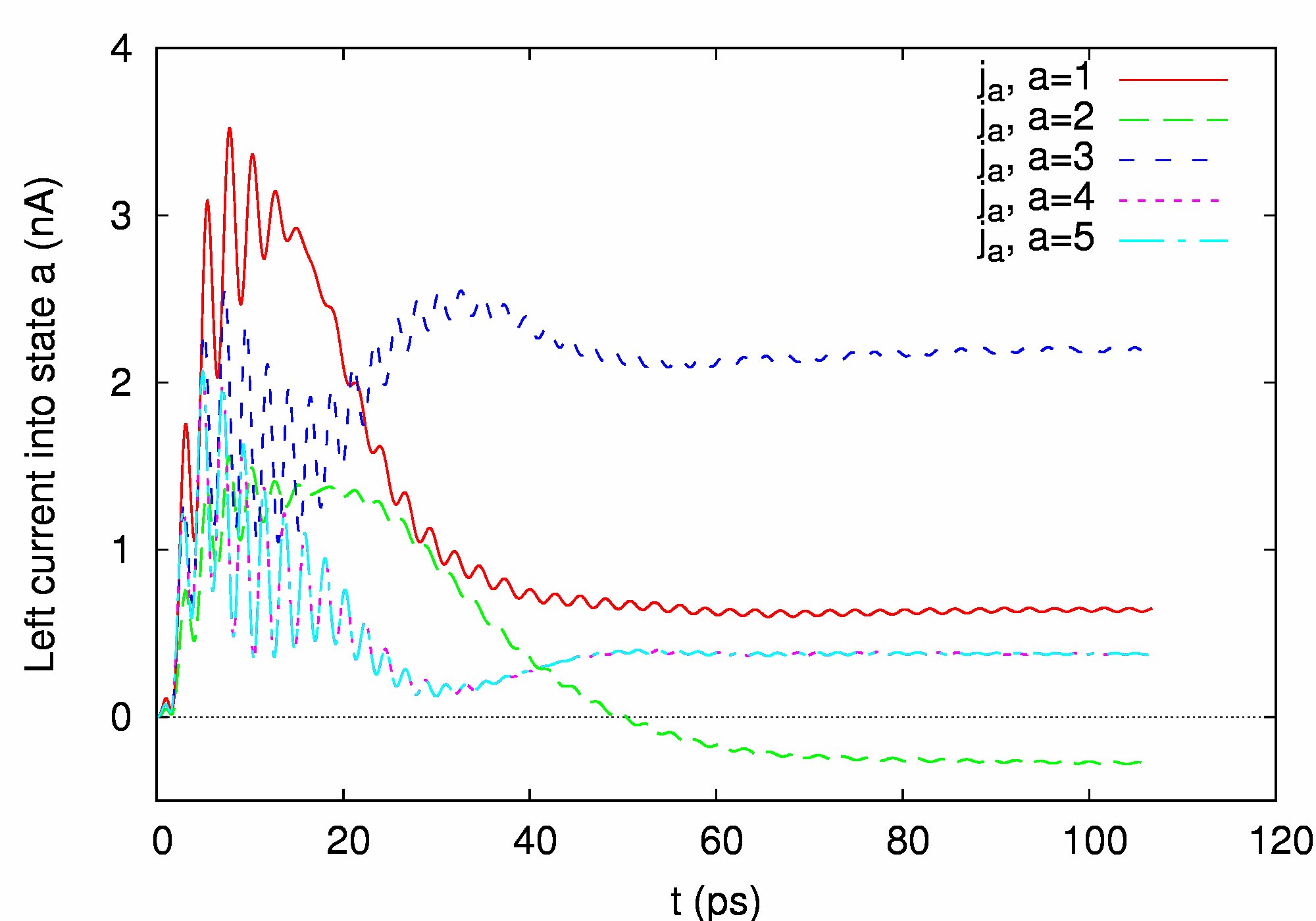}
      \includegraphics[width=0.49\textwidth,angle=0]{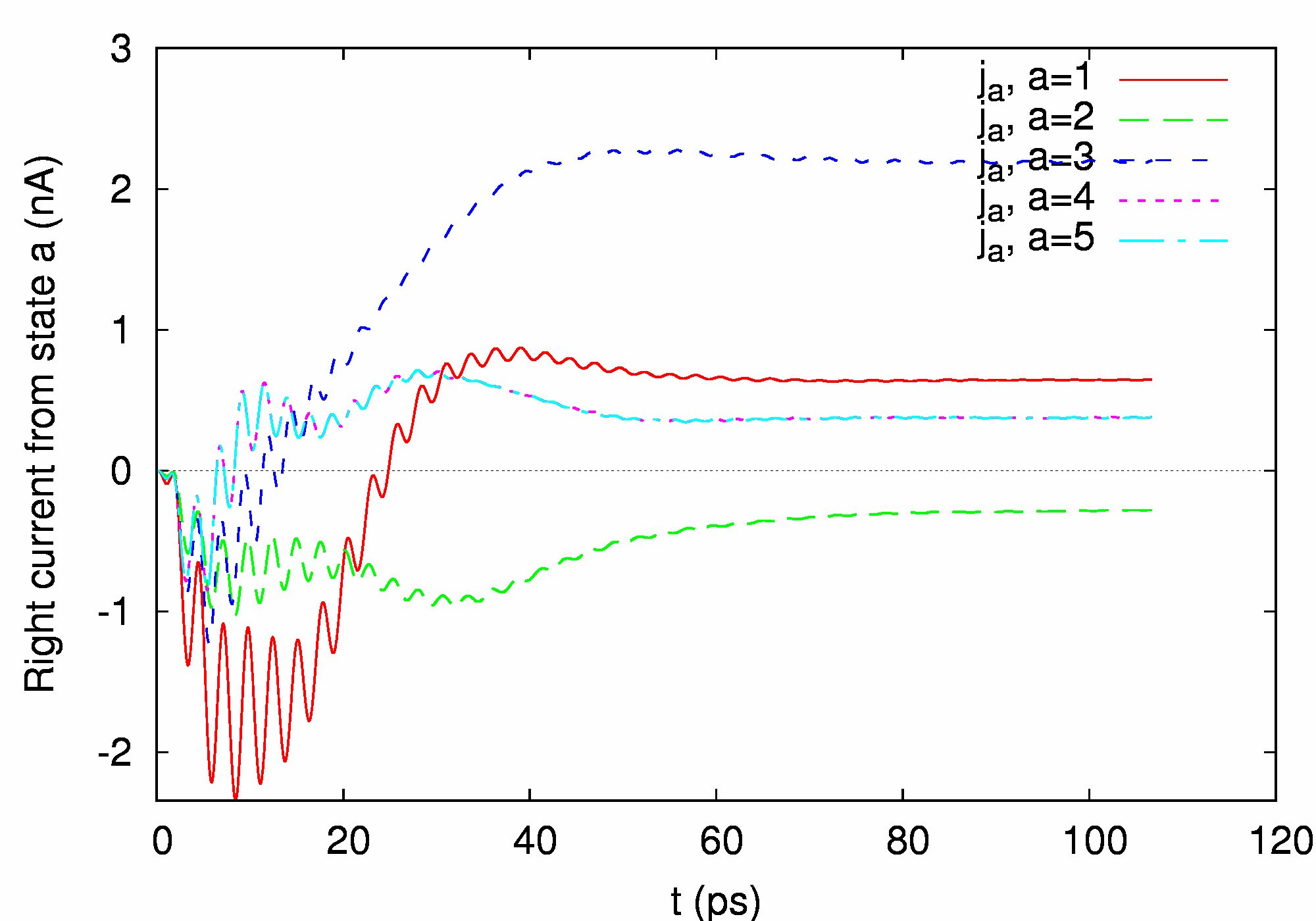}
      \end{center}
      \caption{The current from the left lead into the SES
               $a$ (left panel), and the current from the SES $a$ into the right
               lead (right panel) vs.\ time for the system with an embedded off-centered
               Gaussian well. $y_0=60$ nm, $\Delta =0.2$ meV.
               $L_x=900$ nm, $\hbar\Omega_0=1.0$ meV, $V_0=-2.0$ meV,
               $\beta_x=\beta_y=0.03$ nm$^{-1}$, $g_{00}^l=10.0$ meV, and $x_0=0$.}
      \label{LRja_V2m_y0p60_g2}
\end{figure}

\begin{figure}[htbq]
      \begin{center}
      \includegraphics[width=0.49\textwidth,angle=0,viewport=20 38 320 110,clip]{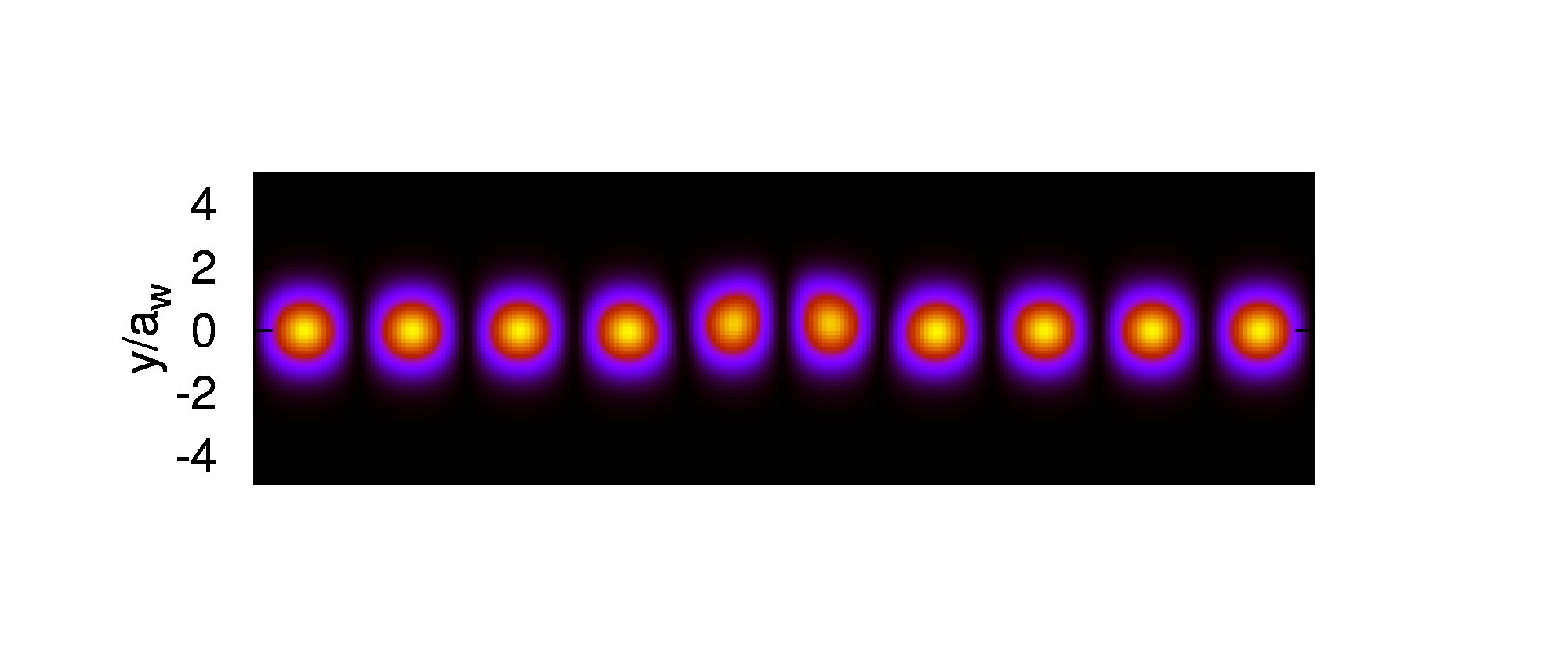}
      \includegraphics[width=0.49\textwidth,angle=0,viewport=20 38 320 110,clip]{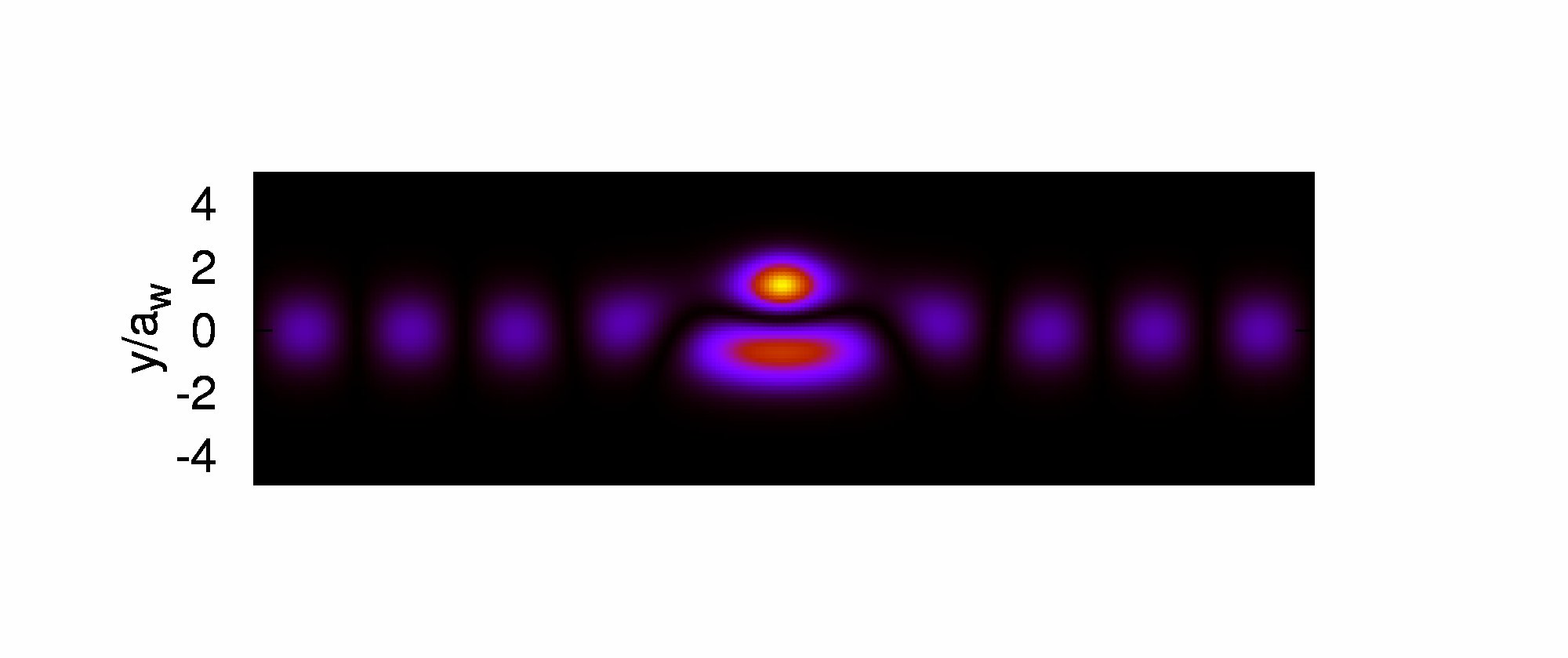}\\
      \includegraphics[width=0.49\textwidth,angle=0,viewport=20 38 320 110,clip]{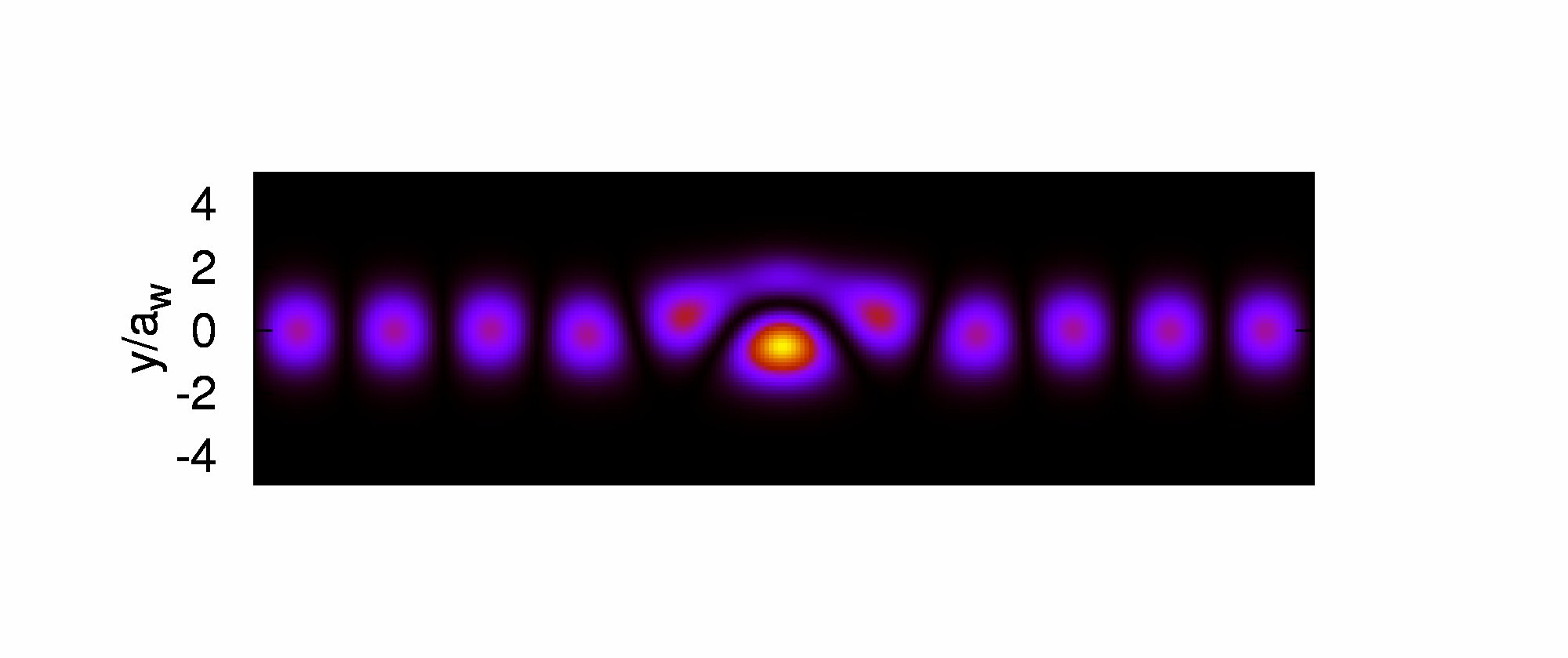}
      \includegraphics[width=0.49\textwidth,angle=0,viewport=20 38 320 110,clip]{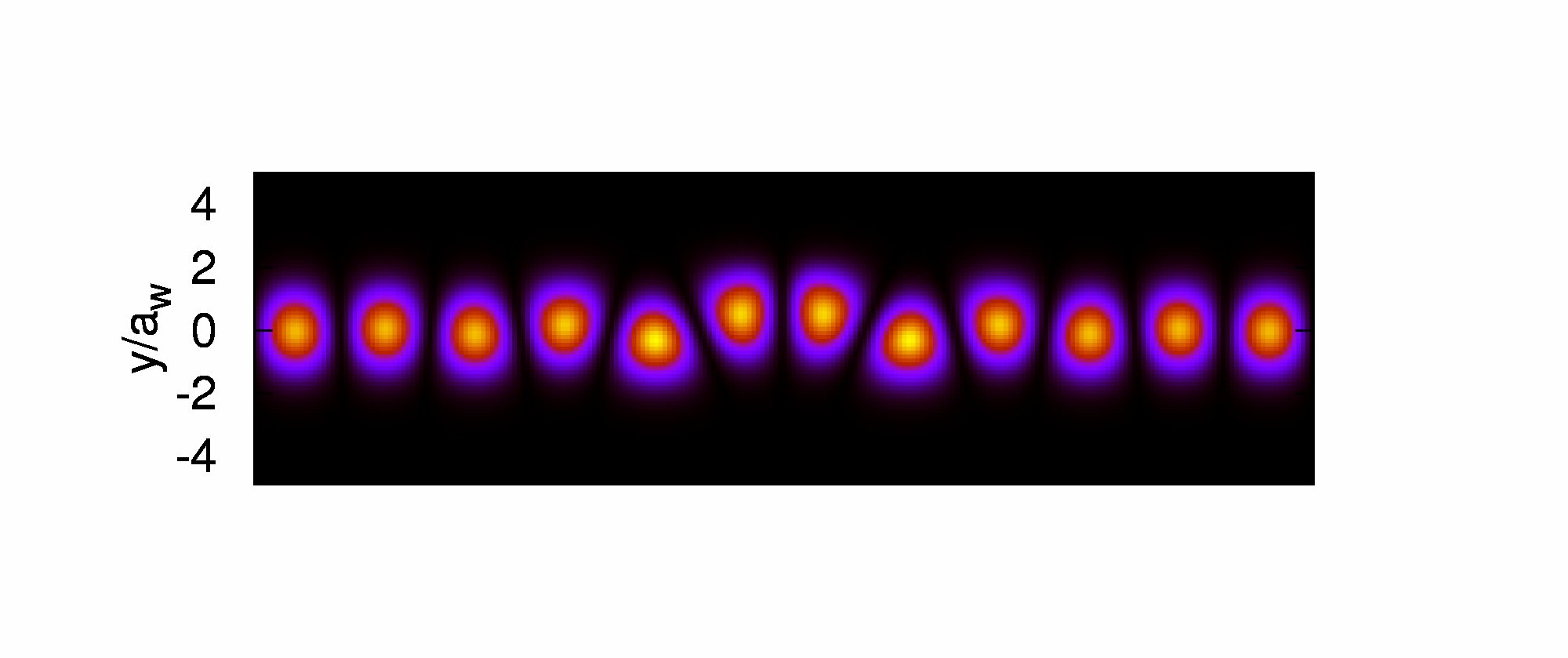}\\
      \includegraphics[width=0.49\textwidth,angle=0,viewport=20 02 320 110,clip]{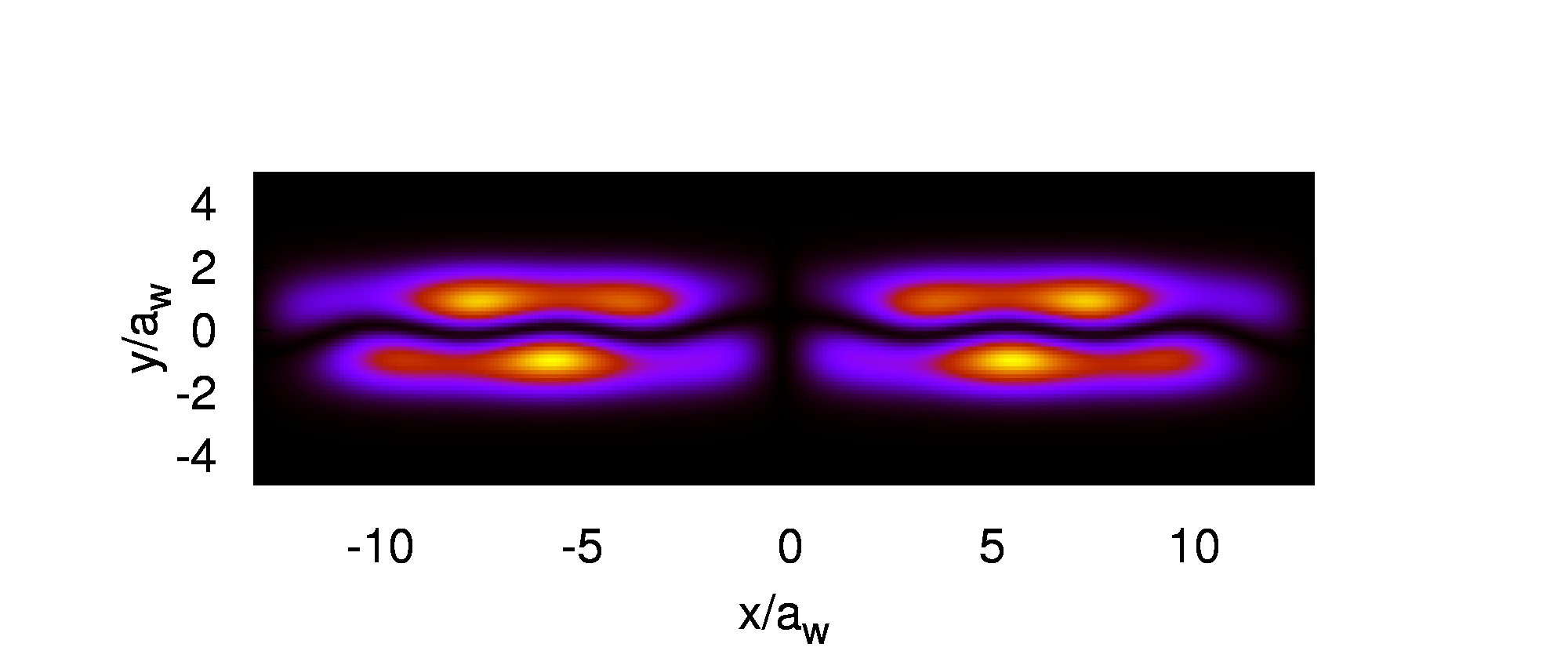}
      \includegraphics[width=0.49\textwidth,angle=0,viewport=20 02 320 110,clip]{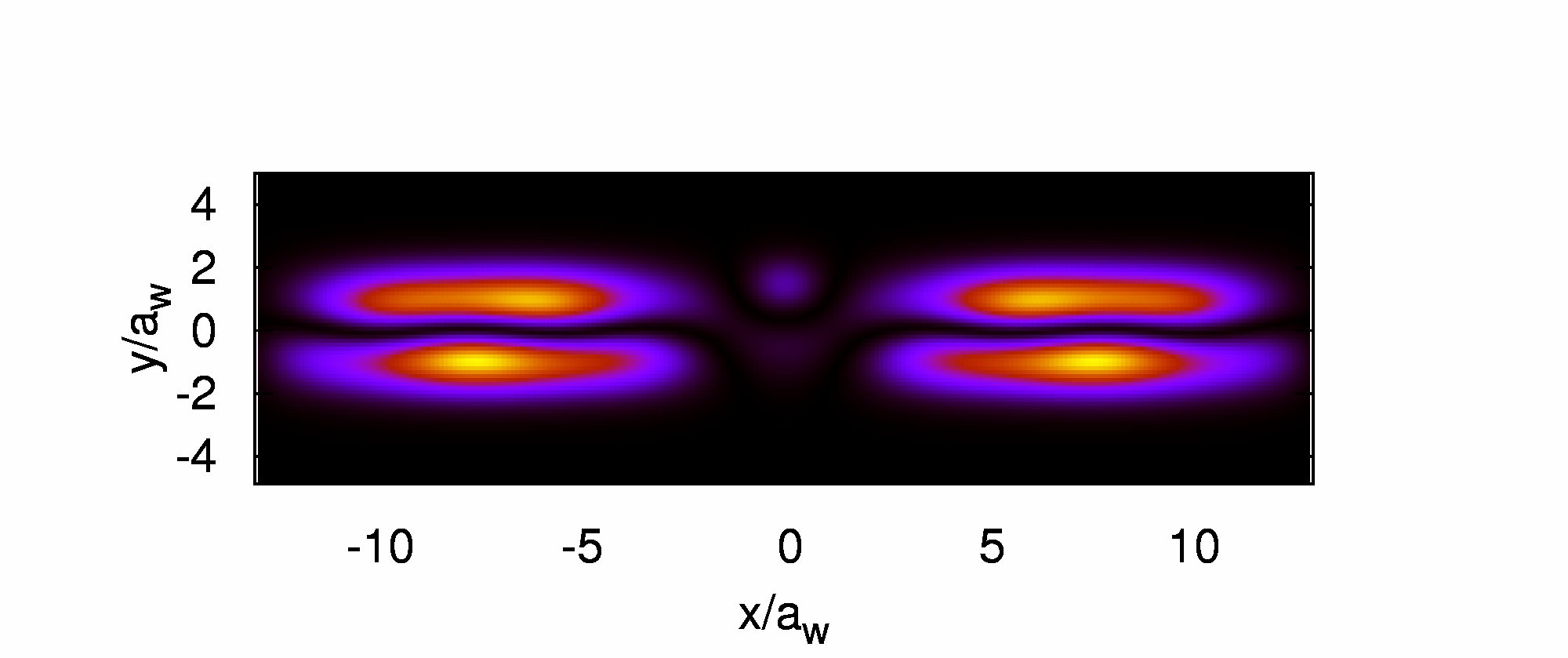}
      \end{center}
      \caption{The probability density of the single-electron eigenstates of the system
               labelled by $a$ in numerical order with $a=1$ at the top left along rows
               to $a=6$ at the
               bottom right. In the system is an  embedded off-centered Gaussian well.
               $L_x=900$ nm, $\hbar\Omega_0=1.0$ meV, $V_0=-2.0$ meV, $y_0=60$ nm,
               $\beta_x=\beta_y=0.03$ nm$^{-1}$, and $x_0=0$.}
      \label{Wf_V2m_y0p60_g2_a}
\end{figure}

In order to investigate further the behavior of the current $j_2$ through the second
relevant SES in the extended bias window (see Fig.\ \ref{Lja_V2m_y0p3060_g15})
we repeat the calculation for a stronger coupling $g_{00}^l=10.0$ meV, and a slightly
wider extension of the bias window by using $\Delta=0.2$ meV. This change of the
window only brings in two extra SES at the top of it. The results are seen in
Fig.\ \ref{LRja_V2m_y0p60_g2}, where we now display both the left and the right
partial currents.
Like expected the occupation of the system here takes a shorter time and the
currents are higher than for the case of lower coupling. There is of course a
slight rearrangement of the individual partial currents, but an eye catching
change is that now the steady state values for $j^\mathrm{L}_2$ and $j^\mathrm{R}_2$
are reversed in comparison with the other partial currents. The partial current
through $a=2$ flows from right to left, but the total steady state current is still
in the expected direction, from left to right. To get an idea why this is happening
we present in Fig.\ \ref{Wf_V2m_y0p60_g2_a} the probabilities of the six relevant SESs.
There we see that the SES $a=2$ is a quasi-bound $p$-state in the off-centered Gauss
well with thus a high probability in the well and just below it, but a much reduced
probability density towards the contact ends of the system. In the well region it has
a character of a state of the second subband, but close to the contacts it has a
character of a state from the first subband. In our earlier calculations with the
Lippmann-Schwinger model of wells in a totally open quantum wire such quasi-bounds
states would always cause a sharp dip in the conductance due to a strong backscattering
\cite{Bardarson04:245308,Gudmundsson05:339,Gudmundsson05:BT}. Here, the off-centering of
the Gaussian well is important in order to enhance the coupling between neighboring
subbands. This explains the fact that the total current thought the system is slightly
lower for the case of the more off-set well, (see Fig.\ \ref{LRja_V2m_y0p60_g2}).
In calculations with the
present GME model we have not found corresponding behavior for an embedded hill in
the system in the case of no external magnetic field. The conclusion is thus close at
hand that the reversal of the partial current here is a manifestation of the total
reflectance of a quasi-bound state in the GME formalism. One might suspect the
fact that the SES $a=2$ is just below $\mu_\mathrm{R}$ playes a role here, but
we have excluded this explanation by shifting the chemical potential window
slightly down to place the state into its center. After this shift the same state
still exhibits a reversed steady state current, so the character of its wave
function plays the main role here and not its location in the energy spectrum
with respect to $\mu_\mathrm{L}$ and
$\mu_\mathrm{R}$.

\begin{figure}[htbq]
      \begin{center}
      \includegraphics[width=0.49\textwidth,angle=0,viewport=20 15 370 210,clip]{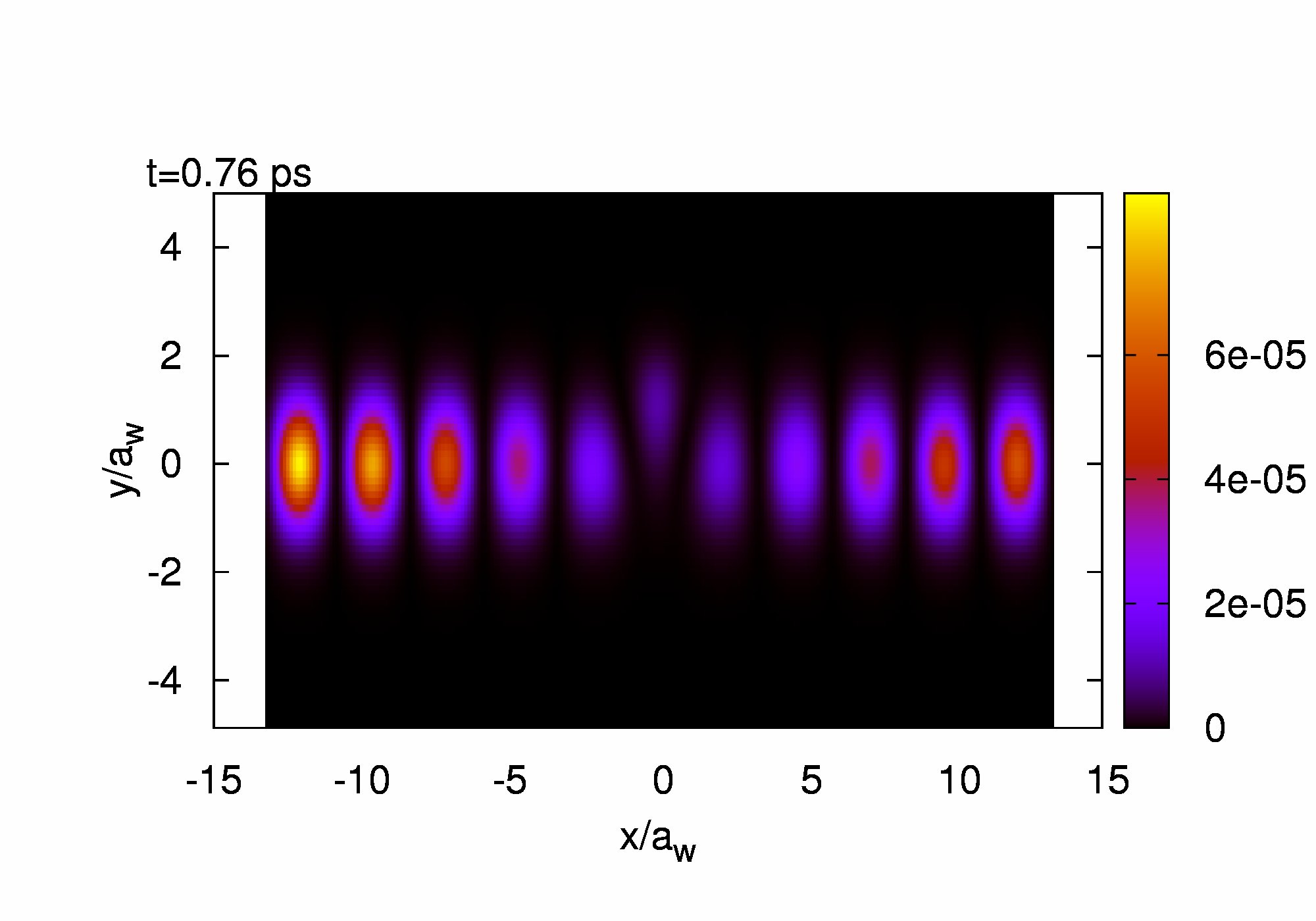}
      \includegraphics[width=0.49\textwidth,angle=0,viewport=20 15 370 210,clip]{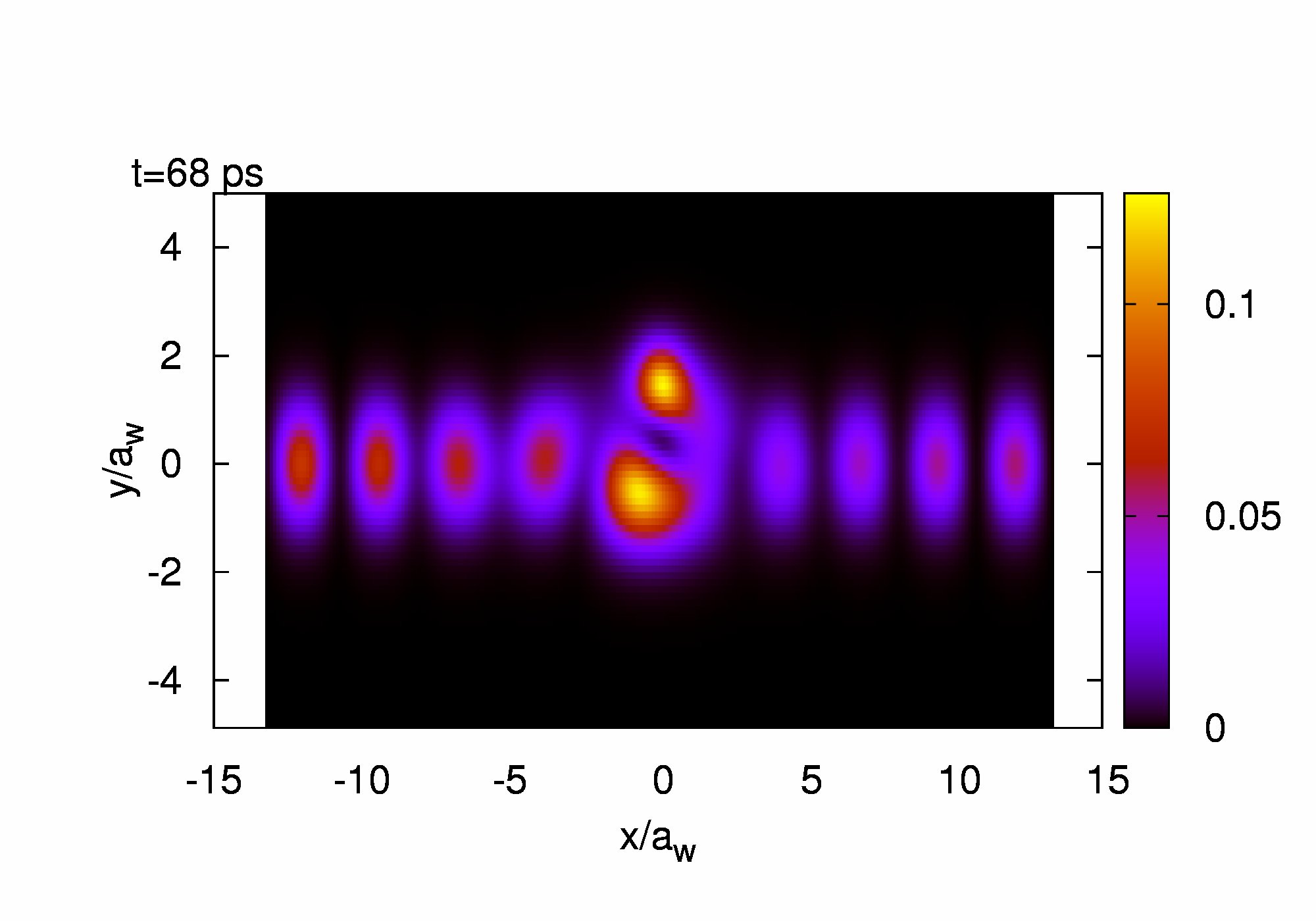}
      \end{center}
      \caption{The average spatial charge distribution for the MES constructed from the
               6 relevant SESs in the extended bias window for $t=0.76$ ps (left panel), and
               $t=68$ ps (right panel) for the system with an embedded off-centered Gaussian well.
               Note the huge difference in scale.
               The system is initially empty $\mu_0=1$.
               $L_x=900$ nm, $\hbar\Omega_0=1.0$ meV, $g_{00}^l=10.0$ meV,
               $\Delta_E^l=1.0$ meV, and $\Delta =0.2$ meV $V_0=-2.0$ meV, $y_0=60$ nm, $x_0=0$,
               and $\beta_x=\beta_y=0.03$ nm$^{-1}$.}
      \label{Qnn_V2m_y0p60_g2}
\end{figure}

\begin{figure}[htbq]
      \begin{center}
      \includegraphics[width=0.49\textwidth,angle=0,viewport=20 15 370 210,clip]{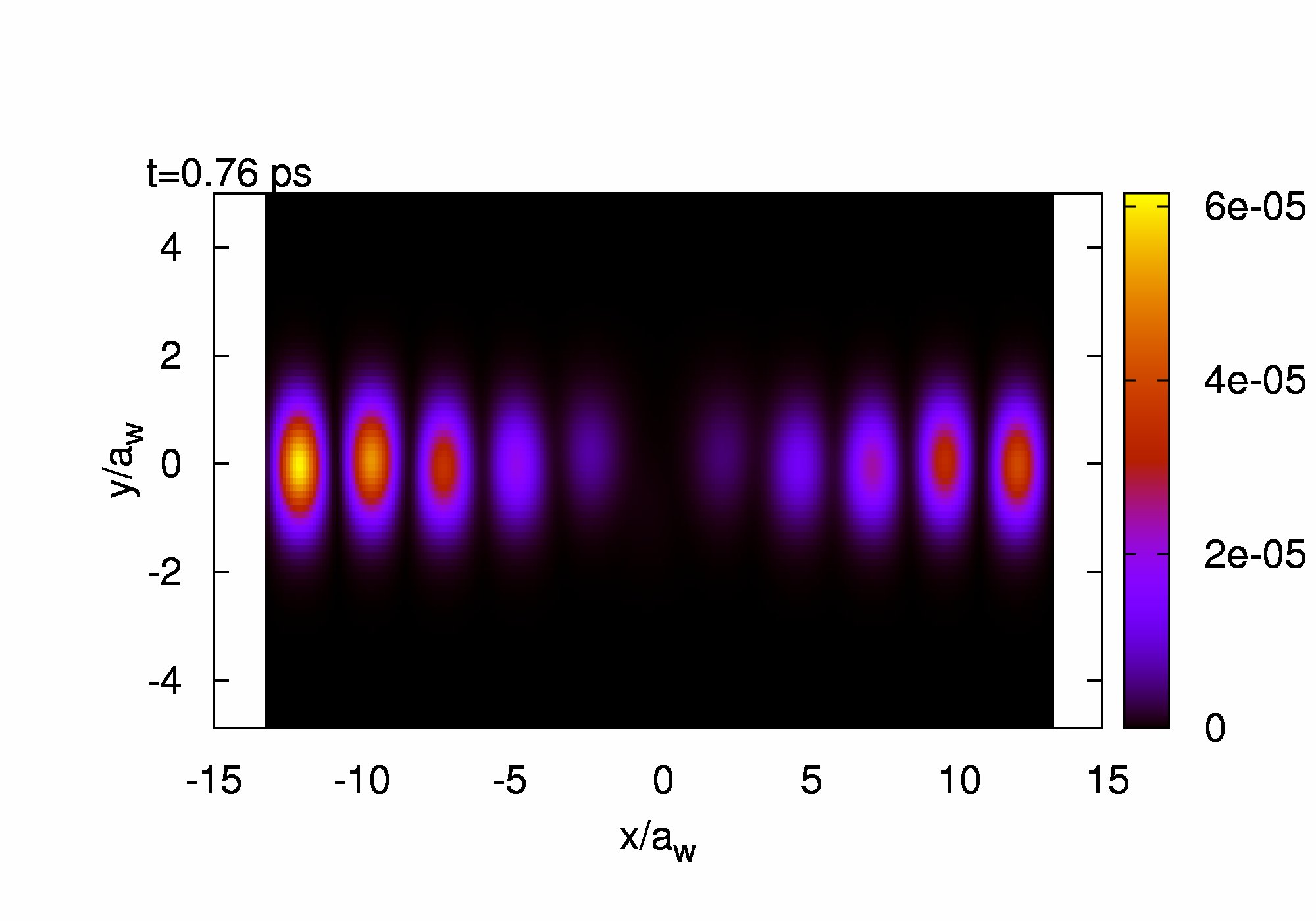}
      \includegraphics[width=0.49\textwidth,angle=0,viewport=20 15 370 210,clip]{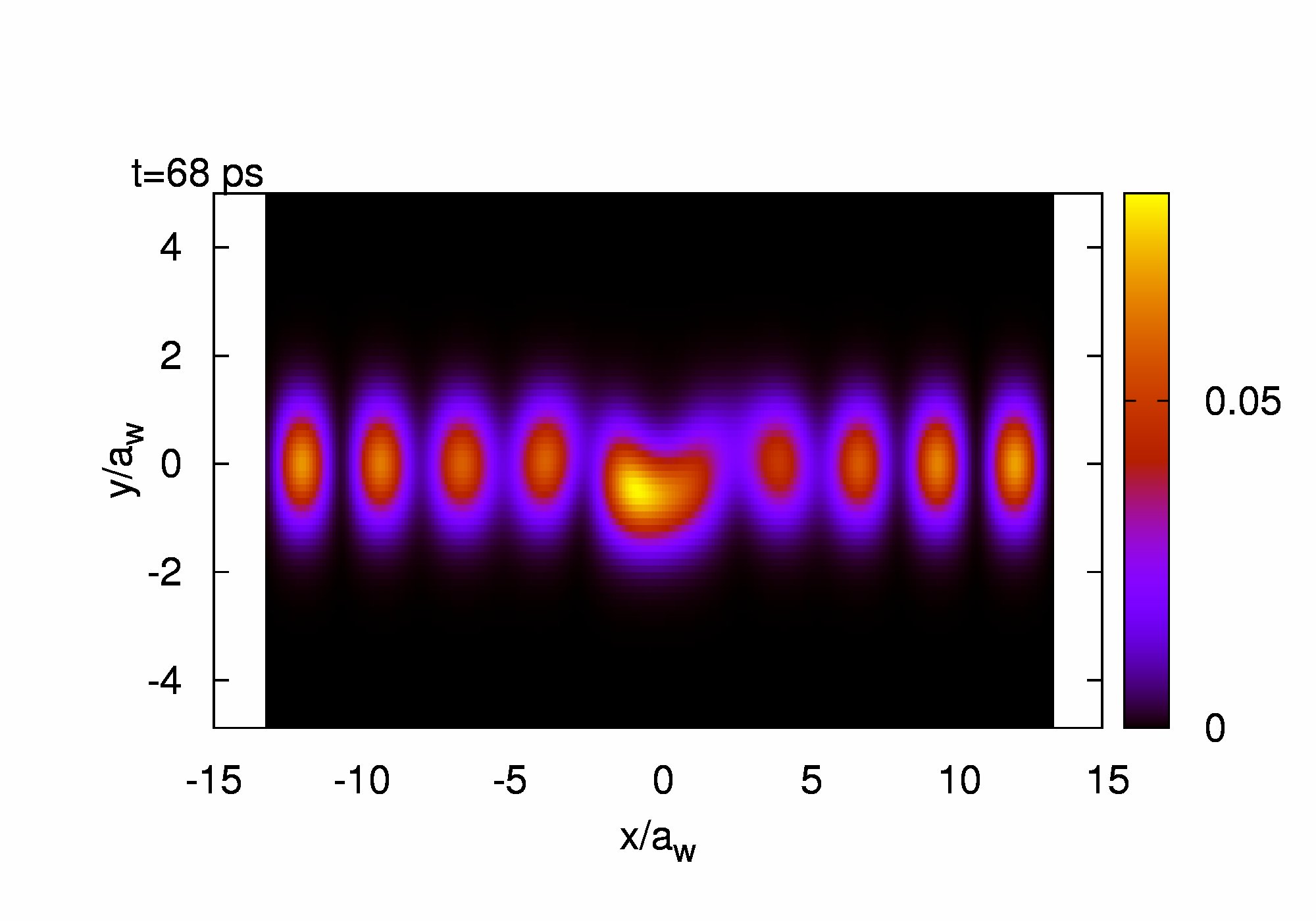}
      \end{center}
      \caption{The average spatial charge distribution for the MES constructed from the
               6 relevant SESs in the extended bias window for $t=0.76$ ps (left panel), and
               $t=68$ ps (right panel) for the system with an embedded off-centered Gaussian hill.
               Note the huge difference in scale.
               The system is initially empty $\mu_0=1$.
               $L_x=900$ nm, $\hbar\Omega_0=1.0$ meV, $g_{00}^l=7.5.0$ meV,
               $\Delta_E^l=1.0$ meV, and $\Delta =0.2$ meV $V_0=2.0$ meV, $y_0=30$ nm, $x_0=0$,
               and $\beta_x=\beta_y=0.03$ nm$^{-1}$.}
      \label{Qnn_V2p_y0p30_g15}
\end{figure}

Another surprise is in store when we look at the time-dependent average spatial charge
distribution for the system
in the beginning and after the system has reached a steady state displayed in
Fig.\ \ref{Qnn_V2m_y0p60_g2}.
At a very early time, $t=0.76$ ps, we see electron probability seeping in from both contact
regions, though more from the higher bias region at the left, and we see already probability
entering the well. In the steady-state regime the system has already entered at $t=68$ ps
we have a strong $p$-state around the well, but the electrons have a higher probability
to be found just outside the off-centered well. This may not be very surprising in light of the
fact that the system carries a good amount of current through it, and we are looking at a
MES here that both carries the information of electrons being quasi-bound in the well, and
being scattered by the well. This latter fact together with the bias difference between the
ends causes the `$p$-state' to be slightly rotated from the $y$-axis.
A further support for this picture can be sought in the results for an off-centered
Gaussian hill shown in Fig.\ \ref{Qnn_V2p_y0p30_g15} for the same two moments of
time explored for the case of the well in Fig.\ \ref{Qnn_V2m_y0p60_g2}.

\begin{figure}[htbq]
      \begin{center}
      \includegraphics[width=0.49\textwidth,angle=0,viewport=20 38 320 110,clip]{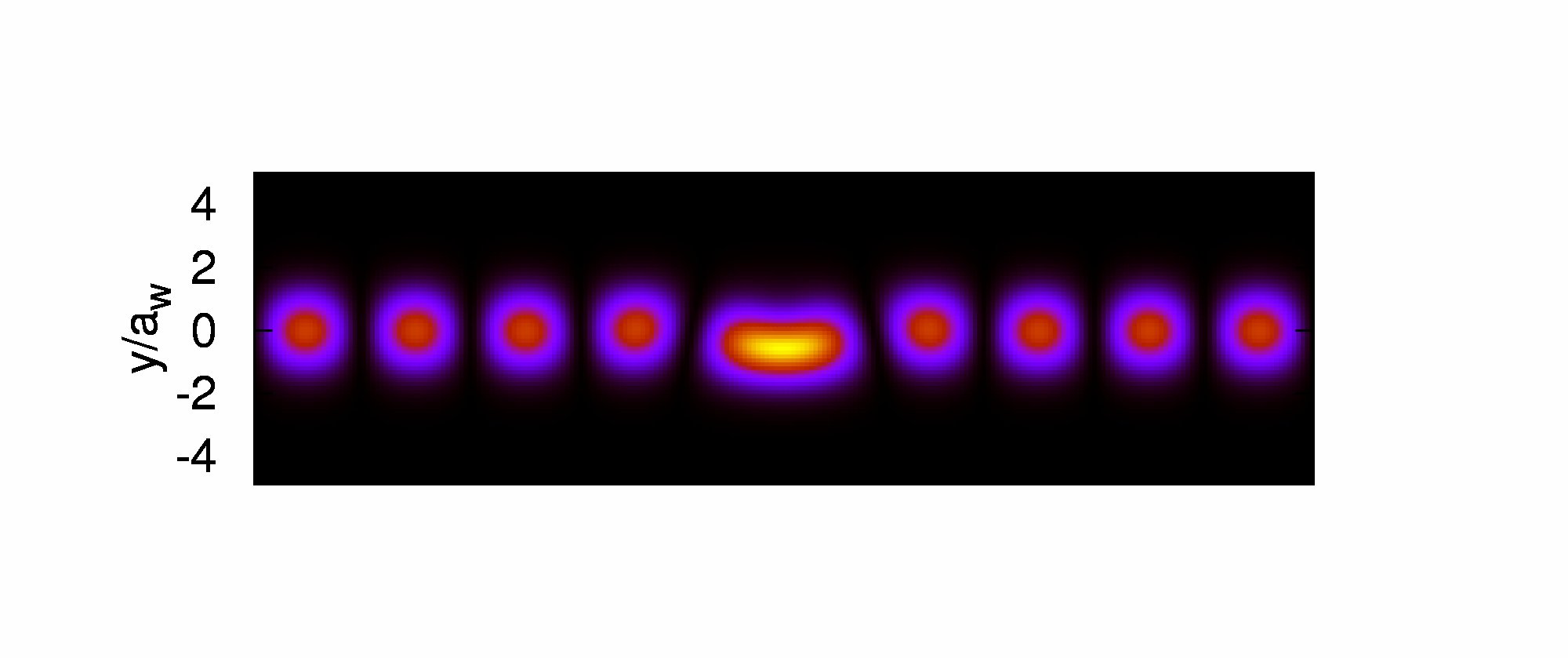}
      \includegraphics[width=0.49\textwidth,angle=0,viewport=20 38 320 110,clip]{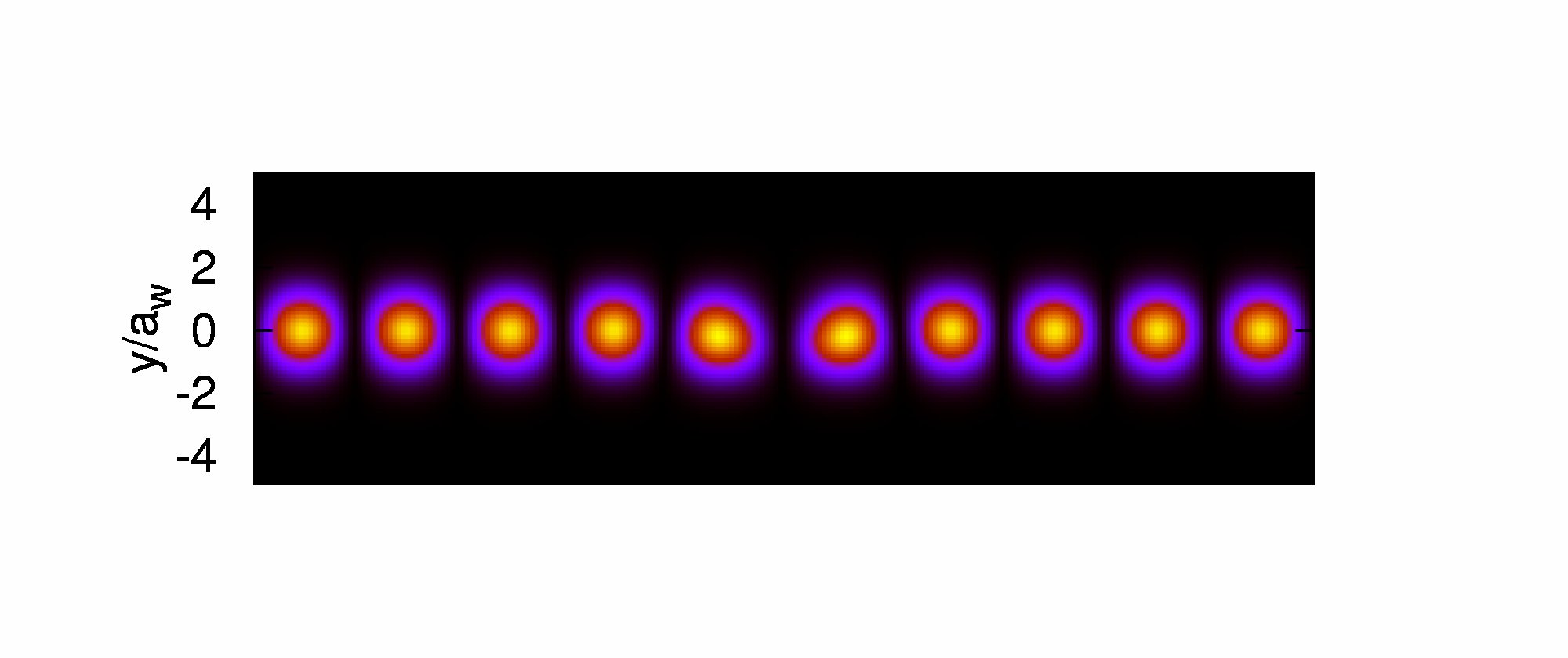}\\
      \includegraphics[width=0.49\textwidth,angle=0,viewport=20 38 320 110,clip]{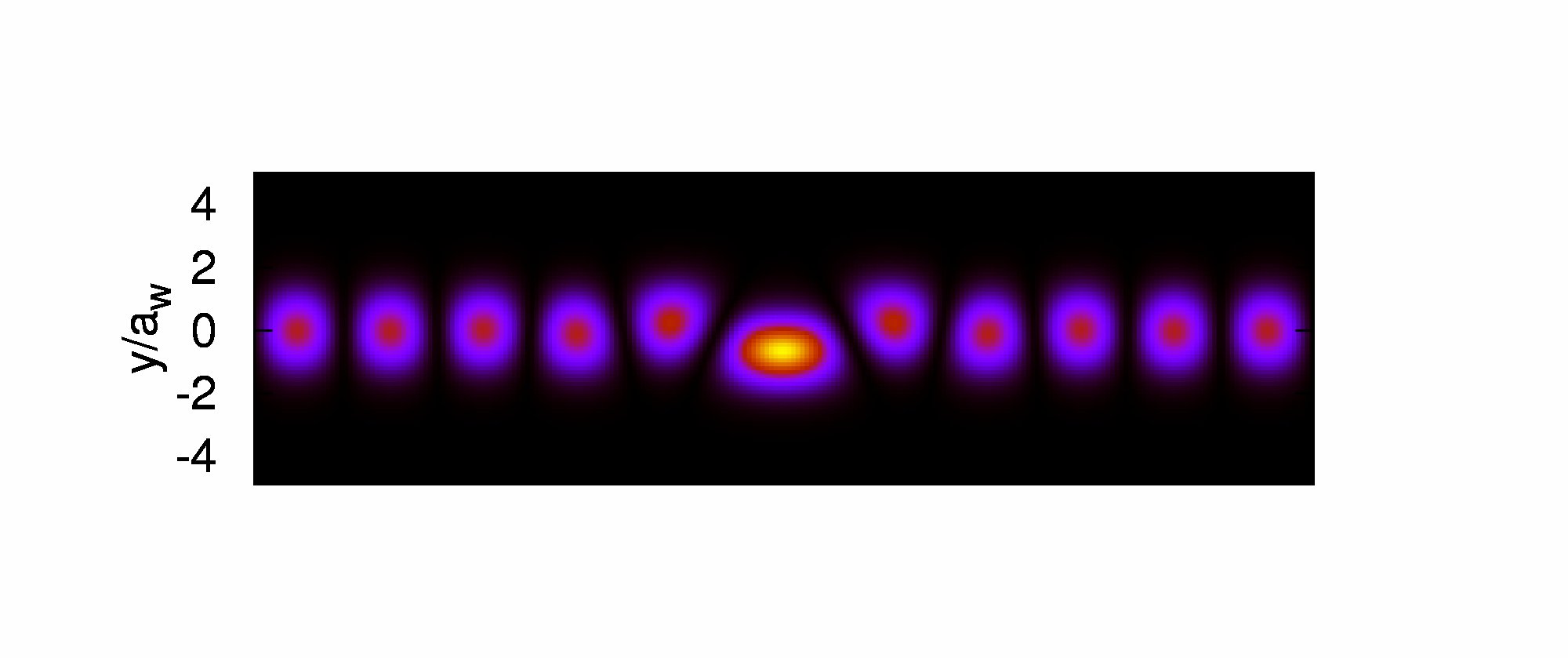}
      \includegraphics[width=0.49\textwidth,angle=0,viewport=20 38 320 110,clip]{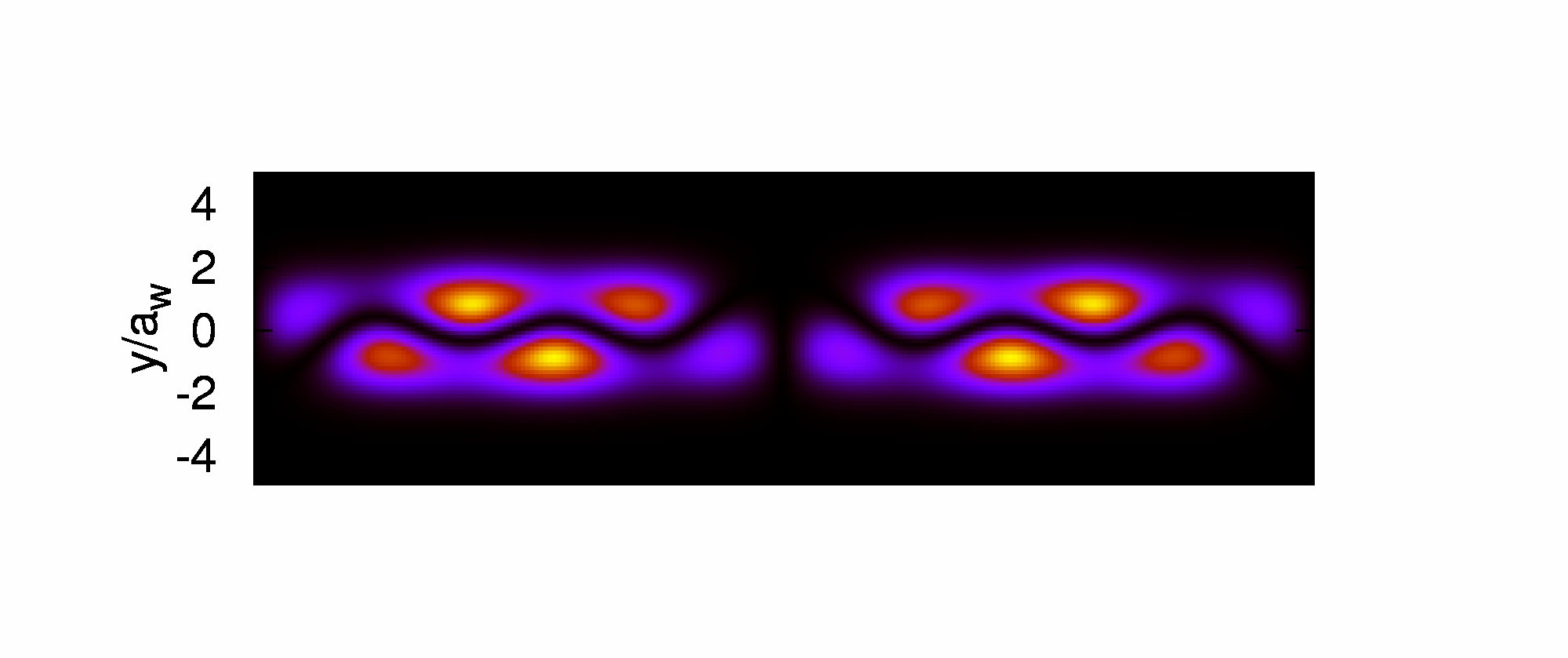}\\
      \includegraphics[width=0.49\textwidth,angle=0,viewport=20 02 320 110,clip]{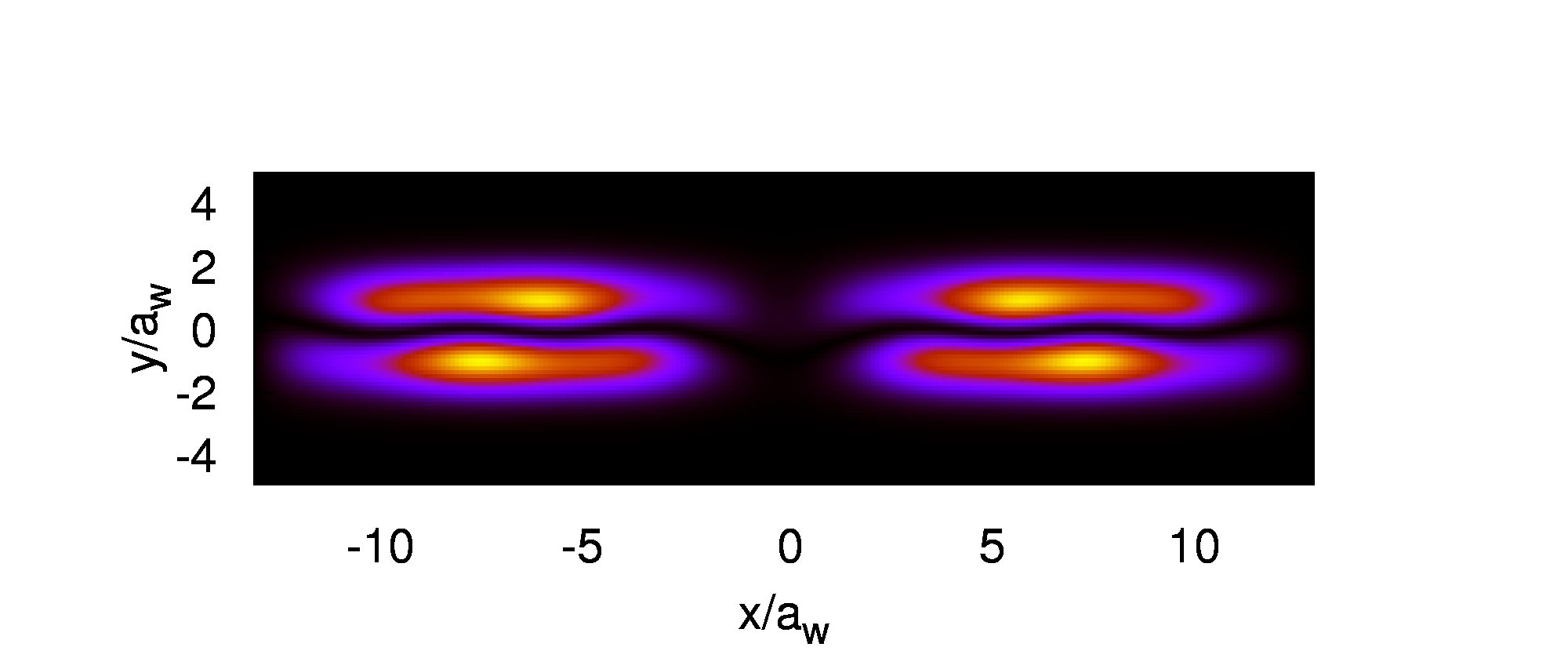}
      \includegraphics[width=0.49\textwidth,angle=0,viewport=20 02 320 110,clip]{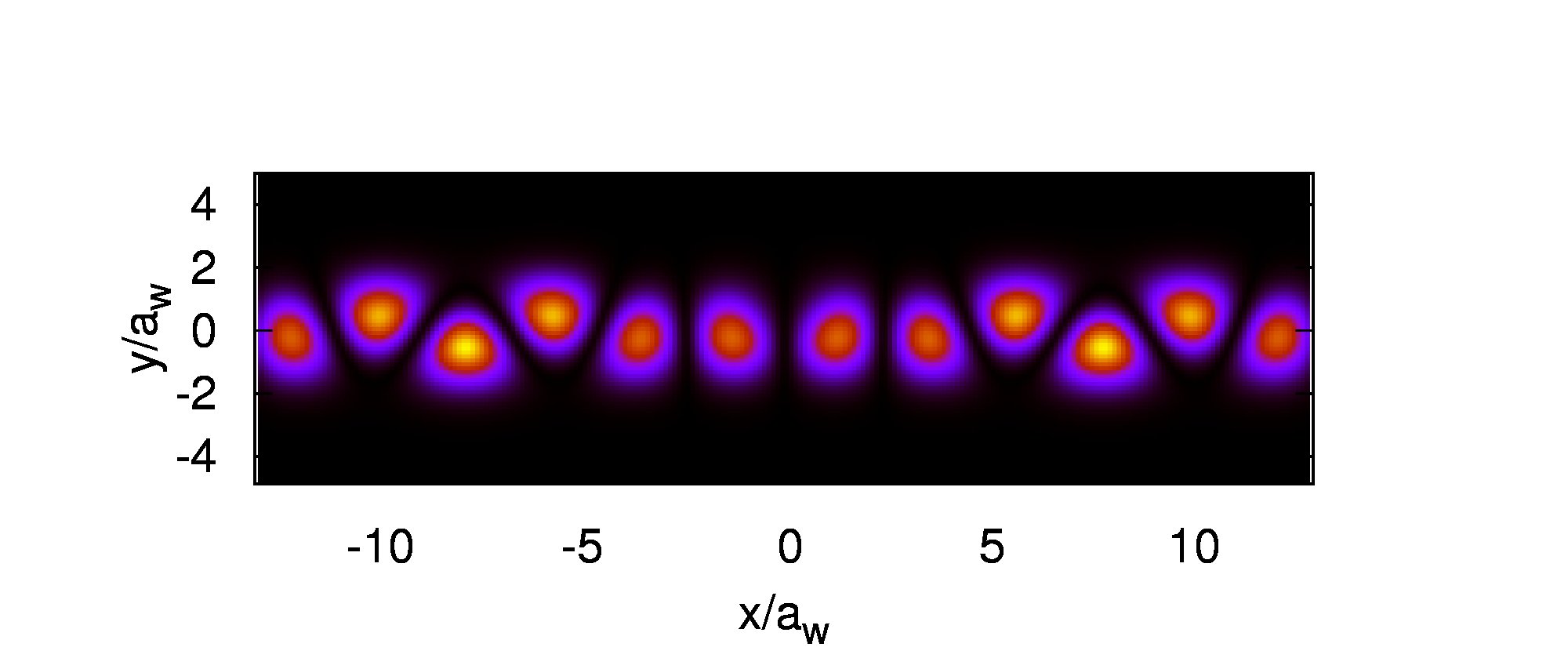}
      \end{center}
      \caption{The probability density of the single-electron eigenstates of the system
               labelled by $a$ in numerical order with $a=1$ at the top left along rows 
               to $a=6$ at the
               bottom right. In the system is an  embedded off-centered Gaussian hill.
               $L_x=900$ nm, $\hbar\Omega_0=1.0$ meV, $V_0=2.0$ meV, $y_0=30$ nm, $x_0=0$,
               and $\beta_x=\beta_y=0.03$ nm$^{-1}$.}
      \label{Wf_V2p_y0p30_g1_a}
\end{figure}

Initially at $t=0.76$ ps we see the charge seeping into the system from both
contact regions as for the case of the well, but now no extra probability is seen
close to the hill, it is a repulsive potential. When the system has reached a steady
state at $t=68$ ps we find out that the electrons have the highest probability
to be found in the system close to the hill.
The Gaussian hill is off-centered, thus mixing up the motion of the electrons along and
perpendicular to the finite quantum wire, the system.
Again, the average charge density is slightly tilted due to the external bias.
Classically speaking we would
say that electrons are scattered between the hill and the parabolic boundary of the
wire opposite to the hill on the side of the wire with negative $y$-coordinate.
The electrons have to go through this symmetry breaking constriction and spend more
time there than elsewhere in the system.
It is interesting to view the probability distribution of the relevant SESs
in Fig.\ \ref{Wf_V2p_y0p30_g1_a} to see that this fact is even found in the
stationary eigenstates of the system.
In addition, we see that states $a=4,5$ with a character mainly reflecting the second subband
and thus with higher energies
suggest a classical analogue of zig-zag motion, and even the state $a=6$ with a clear character of
a SES in the first subband displays this zig-zag motion. A peek back at Fig.\ \ref{Wf_V2m_y0p60_g2_a}
reminds us that this behavior was already present for the system with an embedded well, though not
quite as prominent.

\begin{figure}[htbq]
      \begin{center}
      \includegraphics[width=0.54\textwidth,angle=0]{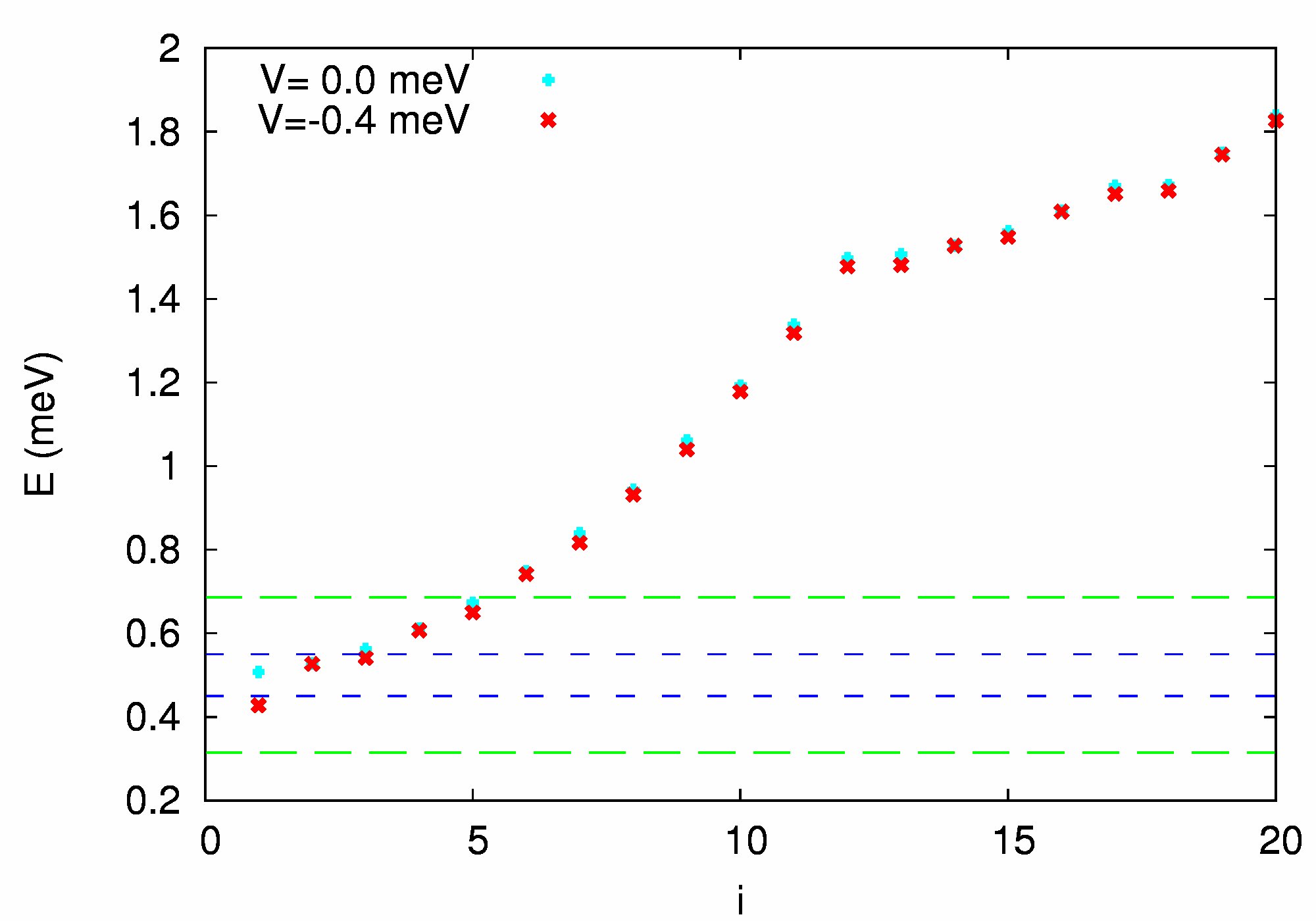}
      \end{center}
      \caption{The energy spectrum of the system (red $\times$)
               vs.\ the SES state number $i$ for the system with an embedded
               centered Gaussian well compared with the spectrum of a
               pure system (blue $+$), and
               the chemical potential in each lead $\mu_L=0.55$ meV, $\mu_R=0.45$ meV,
               and the window of relevant states [$\mu_R-\Delta$, $\mu_L+\Delta$]
               for $\Delta =0.136$ meV. $L_x=900$ nm, $\hbar\Omega_0=1.0$ meV, $V_0=-0.4$ meV,
               and $\beta_x=\beta_y=0.03$ nm$^{-1}$.}
      \label{E_HS_V0p4}
\end{figure}

We have now seen that geometrical properties of the system affecting the bandstructure
in the neighborhood of the bottom of the second subband, where, for example, a
quasi-bound state can be found for the case of an embedded well. What about the
bottom of the first subband? There the states are less coupled to the states in the
leads and carry a smaller amount of a current. In Fig.\ \ref{E_HS_V0p4} we show
the energy spectrum of a system with a centered shallow well and the chemical
potentials selected such that only the bound state of the well is below the
actual bias window.

\begin{figure}[htbq]
      \begin{center}
      \includegraphics[width=0.49\textwidth,angle=0]{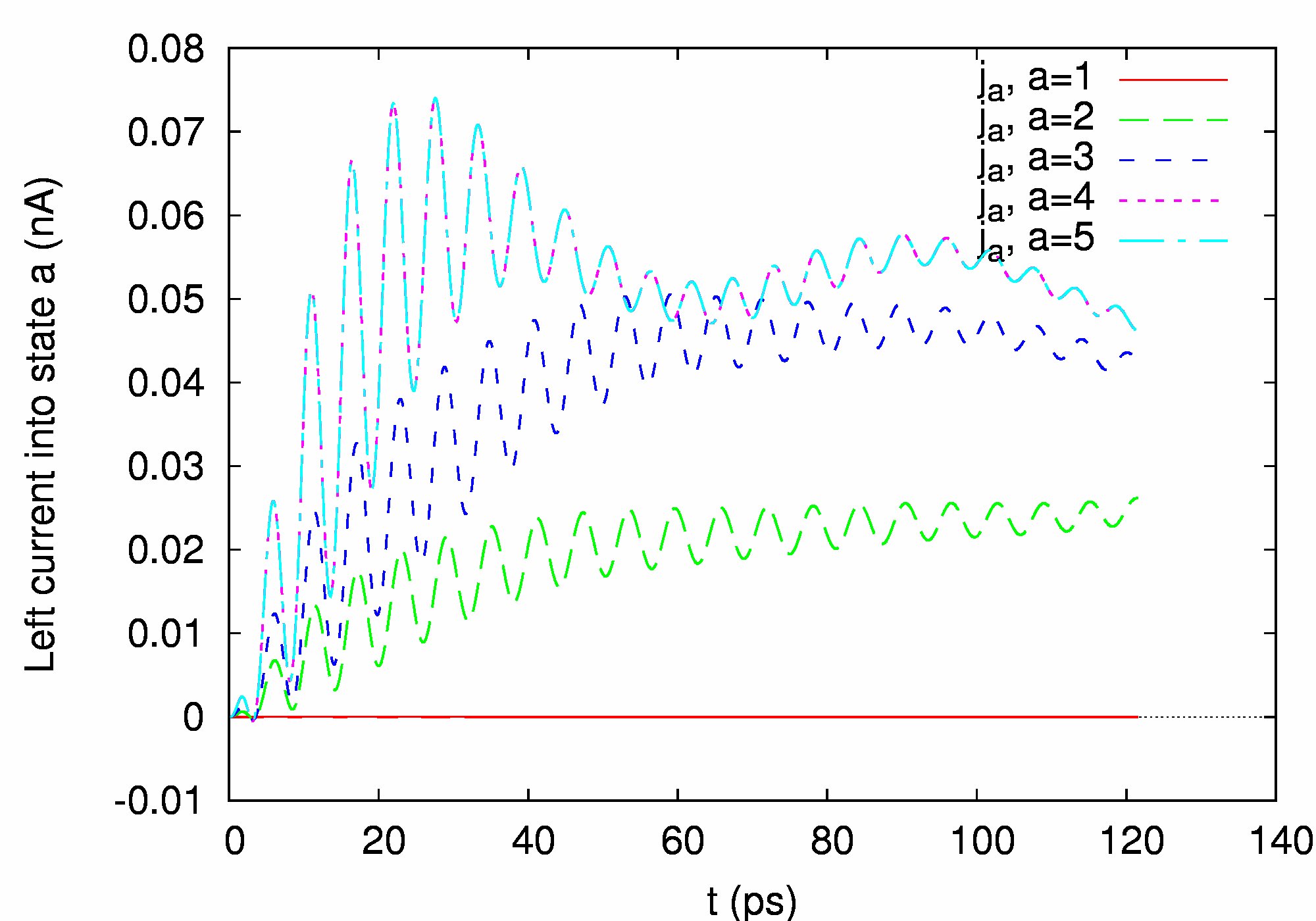}
      \includegraphics[width=0.49\textwidth,angle=0]{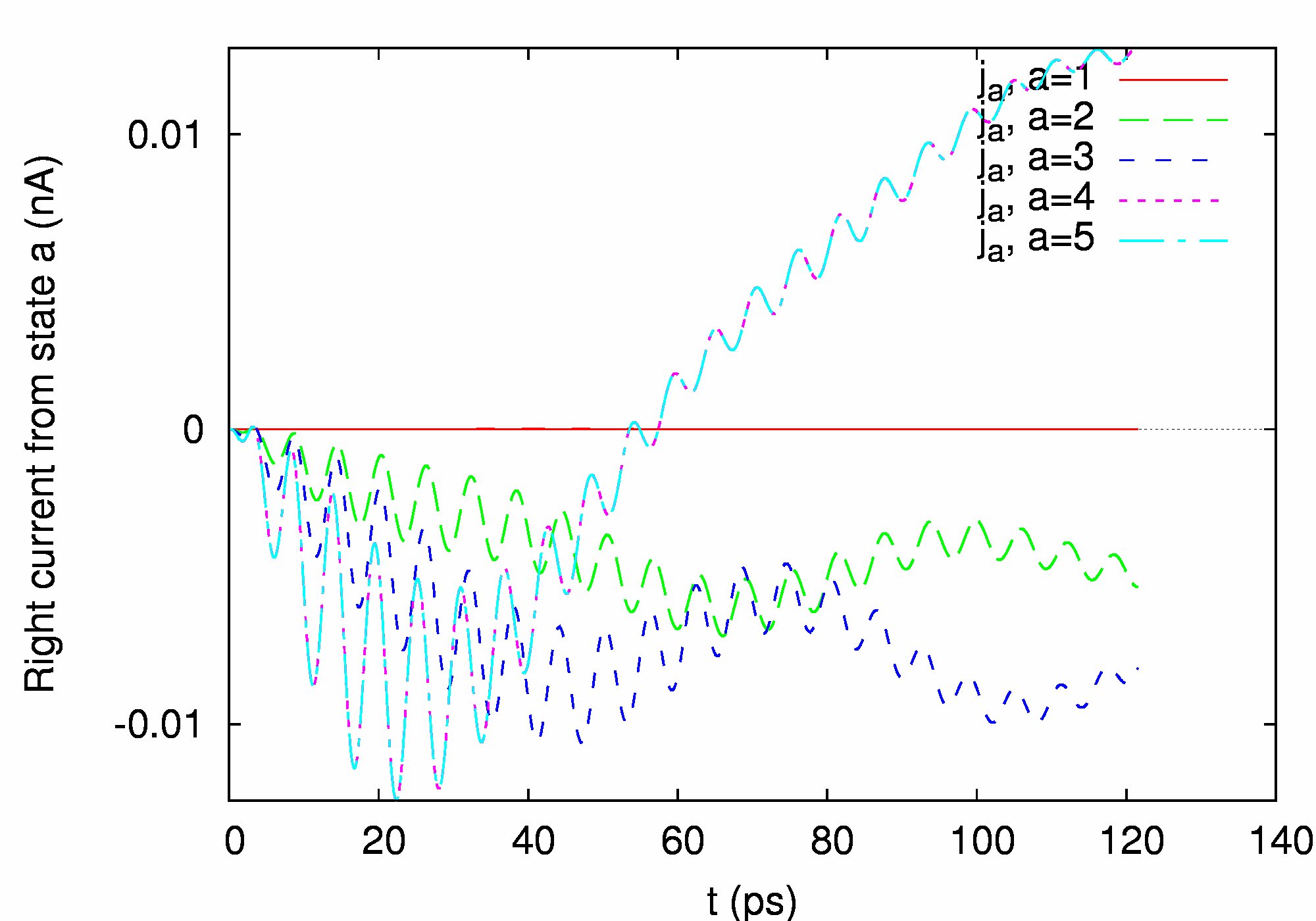}
      \end{center}
      \caption{The current from the left lead into the SES
               $a$ (left panel), and the current from the SES $a$ into the right
               lead (right panel) vs.\ time for the system with an embedded centered
               Gaussian well. $y_0=0$ nm, $\Delta =0.136$ meV.
               $L_x=900$ nm, $\hbar\Omega_0=1.0$ meV, $V_0=-0.4$ meV,
               $\beta_x=\beta_y=0.03$ nm$^{-1}$, and $g_{00}^l=7.5$ meV.}
      \label{LRja_V0p4_g15}
\end{figure}

We do not show the time-dependent occupation of the SESs here, but it would reveal
the fact that even after the $121$ ps the system is far from reaching a steady state, in fact
the occupation of the levels with $a>1$ is still growing linearly.
This can also be verified by observing the partial left and right current for the relevant SESs
in Fig.\ \ref{LRja_V0p4_g15}.
The current through the bound state $j_1$ remains negligible all the time and through
the very small value of the probability for the bound SES $a=1$ in the contact region
a change in its occupation is not expected until on the nanosecond scale. Moreover, the
system is here in the phase that some of the partial current in the right lead are still
directed toward the system, supplying it with electrons.

\begin{figure}[htbq]
      \begin{center}
      \includegraphics[width=0.49\textwidth,angle=0,viewport=20 15 370 210,clip]{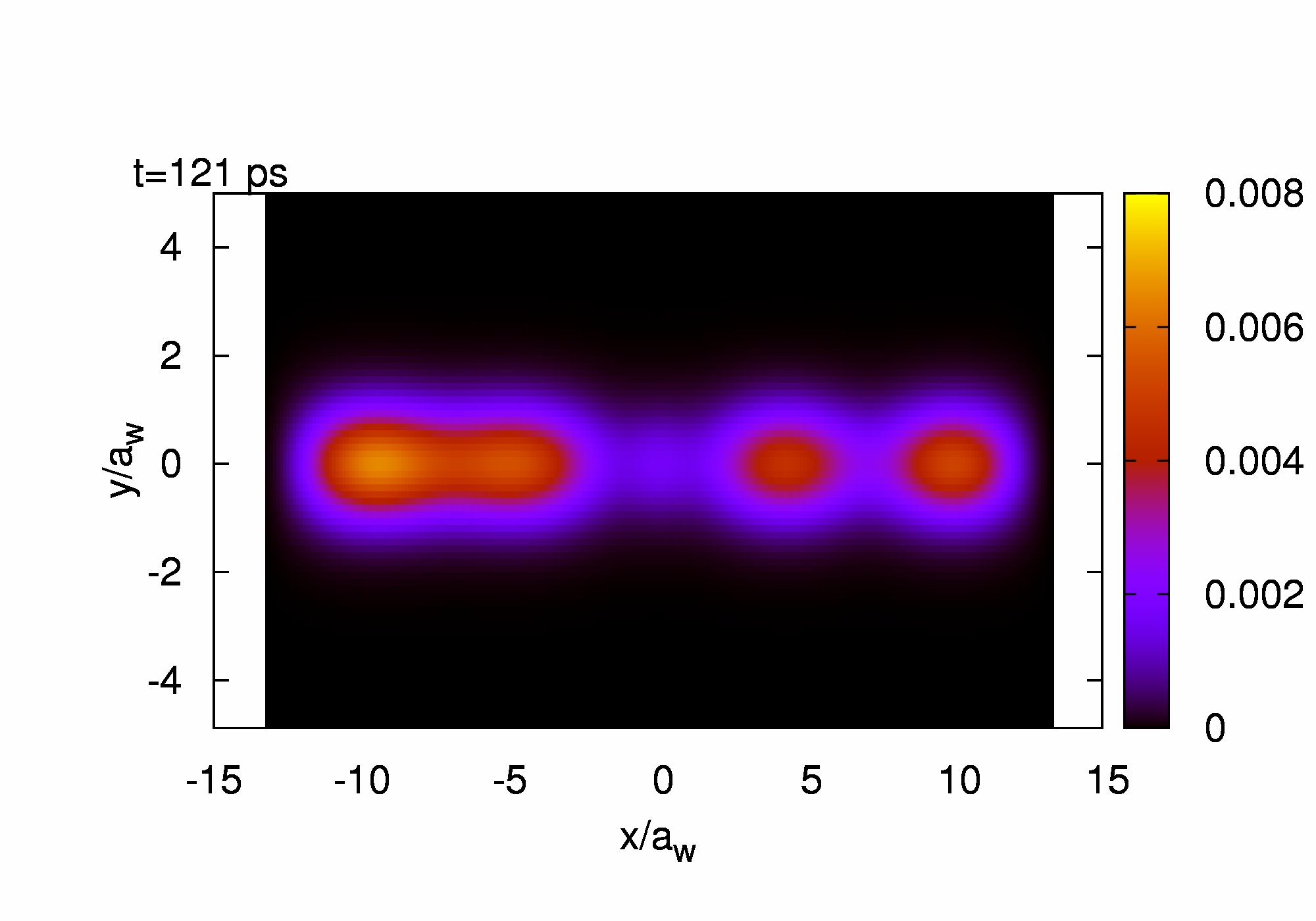}
      \includegraphics[width=0.49\textwidth,angle=0,viewport=20 15 370 210,clip]{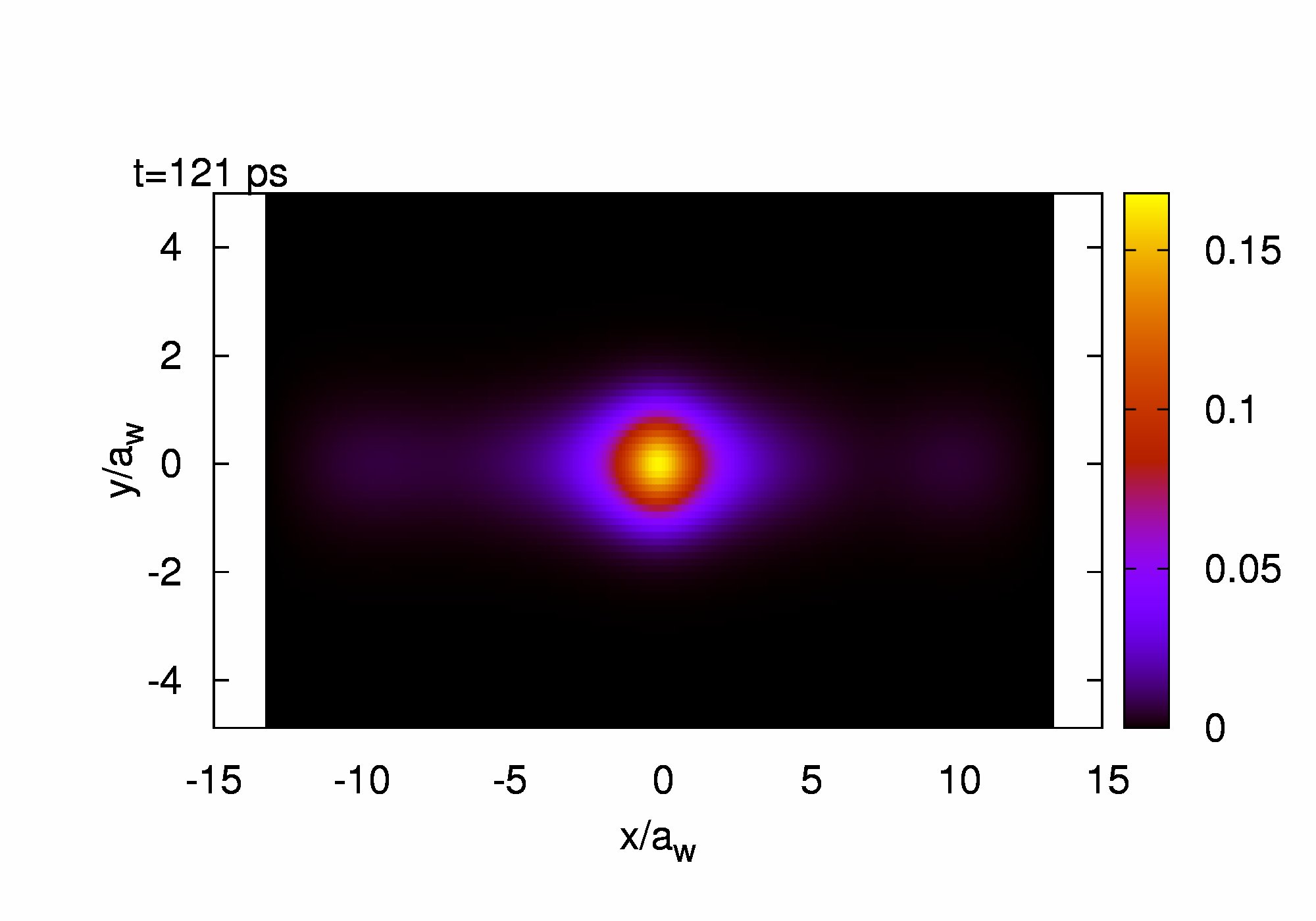}
      \end{center}
      \caption{The average spatial charge distribution at $t=121$ ps
               for the MES constructed from the
               5 relevant SESs in the extended bias window for initially empty system $\mu_0=1$ (left panel),
               and initially occupied by one electron $\mu_0=2$ (right panel) for the system with an
               embedded centered Gaussian well.
               $L_x=900$ nm, $\hbar\Omega_0=1.0$ meV, $g_{00}^l=7.5$ meV,
               $\Delta_E^l=1.0$ meV, and $\Delta =0.136$ meV $V_0=-0.4$ meV, $y_0=0$ nm,
               and $\beta_x=\beta_y=0.03$ nm$^{-1}$.}
      \label{Qnn_V0p4_mu_g15}
\end{figure}

The time-dependent average spatial charge distribution is shown in Fig.\ \ref{Qnn_V0p4_mu_g15}
for $t=121$ ps indicating that the empty system is almost still empty at this time, or in the
case of one electron initially in the system it is still there at this time without any significant
change. The coupling between the bound state and the other states is simply much to small as it
is governed by the behavior of the states in the contact region at the ends of the finite quantum
wire. In contrast, the quasi-bound state corresponding to $a=2$ in Fig.\ \ref{E_HS_V2m_y0p30} 
(the second state from below in the window of relevant states marked by the green dashed lines) had a
significantly higher coupling to other relevant transport states as can be seen in Fig.\ \ref{Wf_V2m_y0p60_g2_a}
from the fact that it has a higher probability in the contact region.
It, was indeed a quasi-bound state, but the SES $a=1$ shown in Fig.\ \ref{Qnn_V0p4_mu_g15}
is a real bound state on the time scale shown (121 ps) with no discernible coupling to the relevant
SESs above it in energy. The current through this state oscillates around zero with amplitude many
orders of magnitude lower than the currents through the neighboring states seen in Fig.\ \ref{LRja_V0p4_g15}.

%
%---------------------------------
%
\section{Summary}
In this publication we have shown that the GME formalism can be used to describe
time-dependent transport through a semiconductor system on the nanoscale with
complex geometry connected to broad leads with several active modes. We have
focused our attention on internal processes in connection with bound and
quasi-bound states in the system, rather than attempting to describe specific
systems of experimental interest. In this way we have been able to compare the
results of the GME formalism to earlier calculations with the Lippmann-Schwinger
formalism \cite{Bardarson04:245308,Gudmundsson05:339,Gudmundsson05:BT}.
The model presented here is very flexible, in the
sense that different parabolic confinement can be chosen for each part, the left and the
right leads, and the central system. The coupling to each lead can
be configured individually in time and space. In addition, the embedded system or even
the shape of the central system can be changed without a problem. All this gives
interesting possibilities that have to be explored in the future work in closer
contact with experimental work.

Geometrical effects like selection rules for processes
between the subbands of the system and the leads have
been successfully incorporated in the model by selecting
a coupling of the type represented by Eq.\ (\ref{T_aq}).
We have found the resulting dynamic current through the
system to depend quantitatively on the ``size'' of the
contact area by varying $\delta_1^l$ and $\delta_2^l$, but
the qualitative behavior of the current is not very
sensitive to variation of these parameters within the
same magntitude.

It is only fair to suggest possible experimental systems or setups at this
stage where our GME model could be tested.
The first direct implimentation could be a long broad quantum wire
with the coupling region, indicated by green or shaded areas in Fig.\ \ref{System},
realized by gate strips situated above the lead-sample contact regions,
controlled by external potentials. Here, like in our model the subband structure
in the leads and the system would have a large influence on the transport
through the system. Chen et al.\ \cite{Chen09:012105}
have measured the static magnetotransport properties
of a quasi-1D quantum wire where one might investigate whether the coupling to
the broad ,,leads`` is better described by our GME model appropriate for
weak coupling or the Lippmann-Schwinger approach appropriate for
a more ,,open`` access or coupling to the leads.
Definitely, a more developed GME model could be appropriate to explore
the geometrical and dynamic properties of the single quantum dot
studied by Astafiev et al \cite{Astafiev02:085315}.

The numerical accuracy of our results presented here has been tested by
including more SESs in the calculation of the states and the energy
spectrum of the system, and by including more subbands in the leads and
denser $q$-points in the integration of the GME. We have carefully
selected the time-mesh fine enough for the GME and where possible we
have experimented with different sizes for the window of relevant states
determined by $\Delta$.

We are aware that the time-dependent properties of the coupling have to be selected
in accordance with the relatively narrow energy range of the MESs constructed for
the central system. This set of MESs may only be appropriate for time-dependence
that is not too fast or strong with respect to the window of relevant SESs.

Moreover, it is clear that electron-electron interaction effects neglected here may be of paramount
interest in experimentally relevant systems. Our only excuse is that we have
here taken the first steps to use the GME formalism for a system with rich
geometry without resorting to the Markov approximation. Inclusion of the Coulomb
interaction is in no way trivial. We are dealing with a system with a variable
number of electrons where different charging regimes may be of importance
depending on the type of coupling between the leads and the system.
The huge number of MESs needed for the description of the Coulomb interaction 
in order to retain correlation effects (that is, going beyond a mean field approximation)
is a real problem and opens the 
important issue of how to isolate a relevant section of the Fock-space.
This is a nontrivial future task that we will not shy from.

It is well known that the GME used here does not guarantee the reduced density matrix
to be positive definite unless the coupling between then leads and the central system
is sufficiently weak.
We have checked this for each iteration step in our calculation
and show only results here for a positive definite density matrix.
For states high in the subbands negative probabilities may occur earlier
than for states close to the bottom of the first subband, as this phenomenon depends
on the effective coupling of the relevant states to states in the leads.

\ack
      The authors acknowledge financial support from the Research
      and Instruments Funds of the Icelandic State,
      the Research Fund of the University of Iceland, the
      Icelandic Science and Technology Research Programme for
      Postgenomic Biomedicine, Nanoscience and Nanotechnology,
      the Computing Center for Design of Materials and Devices,
      Icelandic Research Fund grant 090025011,
      the National Science Council of Taiwan under contract
      No.\ NSC97-2112-M-239-003-MY3. V.M.\ acknowledges partial financial 
      support from the PNCDI2 programme (grant No.\ 515/2009) and grant No.\ 45N/2009.
      C.S.T.\ is grateful to the computational facilities supported
      by the National Center for High-Performance Computing in Taiwan
      and the University of Iceland. V.G.\ thanks hospitality of
      National United University in Taiwan.
      The authors acknowledge technical
      assistance from Chun-Chia Fan Jiang and discussion with L{\'i}ney
      Halla Kristinsd{\'o}ttir. 
%
%---------------------------------------------
%
\appendix
\section{The single-electron states of the system and the leads}
The SESs of the finite quantum wire and their energy spectrum are found by diagonalizing $h_\mathrm{S}$
(\ref{h_S}) in the basis $\{\varphi_{n_x}\phi_{n_y}\}$ with
\begin{equation}
      \varphi_{n_x}(x)=\left\{ \begin{array}{ll}
                          \sqrt{\frac{2}{L_x}}\cos\left(\frac{n_x\pi x}{L_x}\right), &     n_x=1,3,5,\dots \\
                          \sqrt{\frac{2}{L_x}}\sin\left(\frac{n_x\pi x}{L_x}\right), &     n_x=2,4,6,\dots
                          \end{array}
                  \right. ,
\end{equation}
satisfying the hard wall boundary condition at $x=\pm L_x/2$ and
\begin{equation}
      \phi_{n_y}(y)=\frac{e^{-\frac{y^2}{2a^2_w}}}
      {\sqrt{2^{n_y}\sqrt{\pi}{n_y}!a_w}} H_{n_y} \left(\frac{y}{a_w} \right) .
\end{equation}
The matrix elements of the embedded potential (\ref{V}) have been calculated analytically.

For the semi-infinite leads with the same confinement parameters
we can use the same basis functions for the $y$-direction, but
for the $x$-direction we use
\begin{equation}
       \varphi_q(x)=\frac{1}{\sqrt{2\pi}}\sin{\left[q(x\pm L_x/2)\right]},
\label{phi_q}
\end{equation}
with `$+$' in the left lead and `$-$' in the right lead. $\{\varphi_q\}$ is a complete
orthogonal and $\delta$-normalizable basis for the continuous wavenumber $q\geq 0$.
These eigenfunctions represent the fact that in equilibrium before the system and
the leads are coupled together the states along the leads are standing waves with
an equal amount of left and right propagating waves.

%
%-----------------------------------------------------------------
%
\section*{References}
\bibliographystyle{unsrt}

%
%
%----------------------------------------------------------------------------------------
%
\end{document}